\pdfoutput=1
\documentclass[aps,prd,onecolumn,superscriptaddress,nofootinbib,floatfix]{revtex4-2}
\usepackage{amsmath,amssymb}
\usepackage{graphicx}
\usepackage{xcolor}
\usepackage[colorlinks=true,linkcolor=blue,citecolor=blue,urlcolor=blue]{hyperref}
\graphicspath{{figs/}}
\newcommand{\EllF}{\mathrm{F}}
\newcommand{\EllE}{\mathrm{E}}
\newcommand{\EllK}{\mathrm{K}}
\begin{document}
\title{The relativistic BCS--BEC crossover in closed form:\\ the cutoff as an effective range}
\author{T.~G.~Khunjua}
\affiliation{The University of Georgia, GE-0171 Tbilisi, Georgia}
\date{November 24, 2026}

\begin{abstract}
At zero temperature the mean-field potential of a Dirac fermion with a
pairing gap contains one momentum integral, that of the sum of its two
quasiparticle levels.  We evaluate it in closed form from a geometric
construction: the two levels are distances in a plane, and the momentum line
is a finite arc of a fixed ellipse, along which the charge, the scalar density
and the pair amplitude, summed over the modes, are combinations of Legendre's
three elliptic integrals.  As the upper limit of the momentum grows, the
integral grows by a quartic term, a quadratic term and a logarithm with a
known coefficient.  The two powers are absorbed into a constant and into the
coupling, and with all momenta integrated what remains is a finite combination
of Legendre's integrals and the logarithm of one scale $\lambda$.  The gap
equation and the density of the relativistic BCS--BEC crossover follow in
closed form.  Once the coupling has been traded for the scattering length,
$\lambda$ takes the place of the cutoff that the literature keeps as a third
parameter.  We show that it fixes the effective range of the channel of the
pair: for a pair of Dirac fermions of mass $m$ with the pairing vertex
$\gamma_5$, $r_e=-(4/\pi m)[\ln(2\lambda/m)-1]$.  At low density and a fixed
$k_Fr_e$ the crossover is that of the two-channel model of a Feshbach
resonance with the same scattering length and effective range.  In the
Nambu--Jona-Lasinio model the pion decay constant fixes $\lambda$, the
effective range of the pion channel is $-8\pi f_\pi^{2}/(N_cM^{3})$, and the
constituent mass $M$ of the quark is the one parameter that remains free.
\end{abstract}
\maketitle

\section{Introduction}\label{sec:intro}

A gas of fermions with a weak attraction is a BCS superfluid of large,
overlapping pairs.  With a strong attraction the fermions bind into molecules,
and the superfluid is the Bose--Einstein condensate of the molecules.  One
mean-field ground state connects the two
regimes~\cite{Eagles1969,Leggett1980,NozieresSchmittRink1985}, and the passage
from one to the other is a crossover.  For nonrelativistic fermions with a
contact attraction it has one parameter at zero temperature, the scattering
length in units of the distance between the particles~\cite{Leggett1980}, and
its gap equation and density are known in closed form: in complete elliptic
integrals~\cite{MariniPistolesiStrinati1998} and, independently, in Legendre
functions~\cite{PapenbrockBertsch1999}.

Dense quark matter brings the same crossover to relativistic fermions.  Quarks
pair in a colour superconductor~\cite{BailinLove1984}, and at a nonzero
isospin chemical potential a quark and an antiquark condense as charged
pions~\cite{SonStephanov2001}; in both cases the pairing can be followed from
a BCS state to a condensate of bound pairs.  The relativistic
crossover was studied by Nishida and Abuki~\cite{NishidaAbuki2005}, with dense
quark matter in mind, and has since been studied in models with a contact
interaction~\cite{Abuki2007,HeZhuang2007,SunHeZhuang2007,Brauner2008,He2010,HeMaoZhuang2013}
and in a boson--fermion model~\cite{DengSchmittWang2007}.
It differs from the nonrelativistic one in two ways.  A Dirac fermion has two
levels at each momentum, one for the particle and one for the antiparticle, and
the mean-field potential integrates their sum.  This integral is not one of
the complete elliptic integrals of the nonrelativistic problem, and it has
been evaluated numerically.  The coupling, moreover, no longer absorbs the
whole dependence on the cutoff: after it has been traded for the scattering
length a logarithm of the cutoff remains~\cite{Abuki2007,Brauner2008}, and the
cutoff stays in the equations as a third parameter, next to the scattering
length and the density~\cite{HeZhuang2007}.

In this paper we derive the closed form of the integral from a geometric
figure formed by the two levels, and identify the third parameter.  We keep an
upper limit of the momentum at first, to see how the integral grows with it:
by a quartic term, a quadratic term and a logarithm, whose coefficient is a
polynomial in the mass, the chemical potential and the gap.  The two powers
are absorbed into a constant and into the coupling.  With all momenta
integrated, what remains is a finite combination of Legendre's three elliptic
integrals and the logarithm of one scale $\lambda$, Eq.~\eqref{eq:main}.  The
gap equation and the density of the crossover are derivatives of this one
formula, and solving the crossover means finding the common root of two closed
expressions.
The scale $\lambda$ is the third parameter, and it fixes an effective range: for
a pair of Dirac fermions of mass $m$ with the pairing vertex $\gamma_5$ the
channel of the pair has $r_e=-(4/\pi m)[\ln(2\lambda/m)-1]$, negative and
logarithmic in the scale.  At low density, with $k_Fr_e$ kept fixed, the
relativistic crossover is then that of the two-channel model of a Feshbach
resonance with the same scattering length and effective
range~\cite{GurarieRadzihovsky2007}.  No second channel is put in:
its part is played by the pair itself at momenta far above the mass.  To our
knowledge this reading of the cutoff has not been given before.

In the figure the two levels are the distances from one point in a plane to
two others, their sum is the diameter of a circle whose heights above two
points of the axis obey one relation, and as the momentum grows a point runs
along a finite arc of a fixed ellipse.  On this arc the charge, the scalar
density and the pair amplitude, summed over the modes, are combinations of
Legendre's three integrals, and everything that grows comes from the end of
the arc.
In one space dimension the same two levels are those of the Gross--Neveu
model with pion condensation, where it was noticed that the potential can be
reduced to elliptic integrals~\cite{ZhukovskyKlimenkoKhunjua2010}.
Reference~\cite{companion} obtained the closed form of the same integral in
three dimensions, for the Nambu--Jona-Lasinio model at isospin density, by
another route, Hermite's reduction, and used it to say what a regularization
can change.  The pair of
that model is the pion, and its channel is our example here: the pion decay
constant fixes the scale, the effective range of the channel is
$-8\pi f_\pi^{2}/(N_cM^{3})$, and the mass $M$ of the quark in the vacuum is
the one number that the vacuum leaves free.

The same two levels, under other names for the three parameters, are those of
the quarks of two-colour QCD with a diquark
condensate~\cite{BraunerFukushimaHidaka2009}, of the paired quarks of a
two-flavour colour superconductor~\cite{Shovkovy2005} and of massive Dirac
electrons with $s$-wave pairing~\cite{MizoguchiOgata2015}.  We
work out the crossover at a fixed mass, and apply this reading of the scale to
the pion condensate and to the two-flavour colour superconductor of the
Nambu--Jona-Lasinio model (Sec.~\ref{sec:quark}).

The paper is organized as follows.  Section~\ref{sec:levels} draws the figure
of the two levels.  Section~\ref{sec:arc} takes the integral along the arc and
ends with the main result, Eq.~\eqref{eq:main}.  Section~\ref{sec:crossover}
gives the equations of the crossover with all momenta integrated, and
Sec.~\ref{sec:range} shows that the scale fixes the effective range of the
channel of the pair and takes the limit of low density.
Section~\ref{sec:quark} turns to quark matter in
the Nambu--Jona-Lasinio model: pion condensation, where the pion decay
constant fixes the scale, the two-flavour colour superconductor, and more than
one condensate, where the closed form no longer applies.
Section~\ref{sec:conclusions}
concludes.  The derivations that the
text skips are in the Appendices.  The formulas are checked in the
Supplemental Material~\cite{supp} by the script that generates it
(Appendix~\ref{app:checks}).

\section{Two levels, one figure}\label{sec:levels}

\subsection{The levels and the integral}

We consider a Dirac fermion of mass $m$ at a chemical potential $\mu$, paired
with a partner of opposite charge; $\Delta$ is the gap.  In the basis that
puts the fermion and its partner one above the other, its mean-field
Hamiltonian is
\begin{equation}\label{eq:H}
  H=\bigl(\boldsymbol\alpha\cdot\mathbf p+\beta m-\mu\bigr)\,\tau_3+\Delta\,\tau_1 ,
\end{equation}
where $\boldsymbol\alpha$ and $\beta$ are the Dirac matrices and $\tau_1$,
$\tau_3$ are the Pauli matrices that act between the fermion and its partner.
Between a fermion and an antifermion the pairing term is the vertex
$\gamma_5$ (Sec.~\ref{sec:range}).  The levels of $H$ are $\pm E^{+}$
and $\pm E^{-}$, each doubly degenerate, with
\begin{equation}\label{eq:levels}
  E^{\pm}(p)=\sqrt{(E_p\pm\mu)^{2}+\Delta^{2}},\qquad E_p=\sqrt{p^{2}+m^{2}} .
\end{equation}
Here $E_p$ is the energy of the free fermion.  The chemical potential counts
it from $\pm\mu$, with one sign for the particle and the other for the
antiparticle, and the gap adds in quadrature: $E^{-}$ is the Bogoliubov energy
of a particle near the Fermi level, and $E^{+}$ that of the antiparticle.  The model of
Ref.~\cite{companion} has these levels.  There the fermion is a quark, $\mu$
is half the isospin chemical potential and $\Delta$ is the charged pion
condensate, and the three symbols are $M$, $\nu$ and $\Delta$.

At zero temperature the levels enter the mean-field thermodynamic potential
through the sum of their absolute values, which at each momentum is a multiple
of $E^{+}+E^{-}$.  Every such potential therefore contains, with a factor that
counts the colours, the flavours or the doubling of the basis, the integral
\begin{equation}\label{eq:integral}
  I(\Lambda)=\int_0^{\Lambda}dp\,p^{2}\bigl[E^{+}(p)+E^{-}(p)\bigr] .
\end{equation}
We keep the upper limit $\Lambda$ for a while in order to see how the integral
grows with it; in the end all momenta are integrated.  The integral is not
elementary, because $E_p$ is itself a square root and each level is a root of
a root.

\subsection{Two points and a triangle}

\begin{figure}[t]
  \centering
  \includegraphics[width=0.86\linewidth]{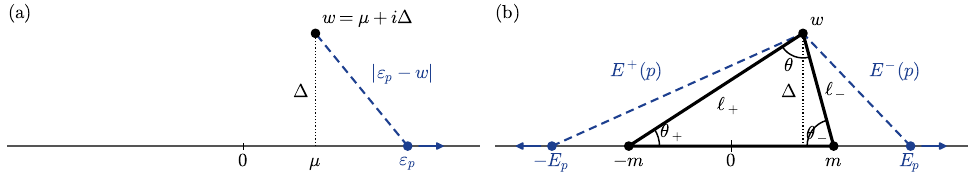}
  \caption{One point and two.  (a)~The level of BCS theory is the distance
  from the fixed point $w=\mu+i\Delta$ to the point $\varepsilon_p$ of the
  real axis, which moves to the right as the momentum grows.  (b)~A Dirac
  particle has two levels, the distances from $w$ to the points $\mp E_p$
  (dashed).  At $p=0$ the points are the ends of the base $[-m,m]$ and the
  levels are the sides $\ell_\pm$ of a triangle; $\theta$ is the angle at the
  apex and $\theta_\pm$ the angle between the base and the side $\ell_\pm$.
  Drawn for $(m,\Delta,\mu)=(1,1.1,0.7)$ and $E_p=1.75$.}
  \label{fig:points}
\end{figure}

The nonrelativistic problem has one level, and that level is a distance.  A
fermion with the kinetic energy $\varepsilon_p$, a chemical potential $\mu$
and a gap $\Delta$ has the level
$\sqrt{(\varepsilon_p-\mu)^{2}+\Delta^{2}}=|\varepsilon_p-w|$ of BCS theory,
with $w=\mu+i\Delta$.  It is the distance from the fixed point $w$ of the
complex plane to the point $\varepsilon_p$ of the real axis, which moves away
from the origin as the momentum grows (Fig.~\ref{fig:points}a).  The
mean-field potential integrates this one distance over the momentum.  In three
dimensions and at zero temperature the resulting integrals are complete
elliptic integrals of the first and second kinds, with which Marini, Pistolesi
and Strinati followed the crossover from BCS pairing to Bose--Einstein
condensation in closed form~\cite{MariniPistolesiStrinati1998}.

A Dirac particle brings a second point.  With the same $w$, the two levels
\eqref{eq:levels} are $E^{-}=|E_p-w|$ and $E^{+}=|-E_p-w|$: the distances
from $w$ to the points $+E_p$ and $-E_p$ of the real axis, one for the
particle and one for the antiparticle (Fig.~\ref{fig:points}b).  At $p=0$ the
two points are $\pm m$, and the levels are the sides of the triangle with base
$[-m,m]$ and apex $w$,
\begin{equation}\label{eq:triangle}
  \ell_\pm\equiv E^{\pm}(0)=\sqrt{(m\pm\mu)^{2}+\Delta^{2}},\qquad
  \cos\theta=\frac{\ell_+^{2}+\ell_-^{2}-4m^{2}}{2\,\ell_+\ell_-},\qquad
  \cos\theta_\pm=\frac{m\pm\mu}{\ell_\pm} ,
\end{equation}
where $\theta$ is the angle at the apex and $\theta_\pm$ the angle between the
base and the side $\ell_\pm$.  As the momentum grows the two points slide
apart along the axis, and the apex stays where it is.

\subsection{The ellipse through the apex}

The potential integrates the sum of the two levels, $S\equiv E^{+}+E^{-}$, and
not each of them.  The squares of the two levels differ by a multiple of $E_p$
and add up to a polynomial in it,
\begin{equation}\label{eq:twoid}
  (E^{+})^{2}-(E^{-})^{2}=4\mu E_p,\qquad
  (E^{+})^{2}+(E^{-})^{2}=2\bigl(E_p^{2}+\mu^{2}+\Delta^{2}\bigr).
\end{equation}
So $S^{2}$ and $(E^{+}-E^{-})^{2}$ have the sum
$4(E_p^{2}+\mu^{2}+\Delta^{2})$ and the product $16\mu^{2}E_p^{2}$: they are
the two roots of
\begin{equation}\label{eq:quad}
  z^{2}-4\bigl(E_p^{2}+\mu^{2}+\Delta^{2}\bigr)z+16\,\mu^{2}E_p^{2}=0,
  \qquad\text{and}\qquad
  E^{+}E^{-}=\frac{S^{2}}{2}-E_p^{2}-\mu^{2}-\Delta^{2} .
\end{equation}
The radicals are gone: the quadratic contains $E_p$ only through its square,
and therefore the momentum only through $p^{2}$.

The points of a plane whose distances to two given points have a given sum
form an ellipse with these points as its foci.  At a given momentum the apex
$w$ therefore lies on the ellipse with foci $\pm E_p$ and major semi-axis
$\tfrac12S$.  Let $h_E$ be its minor semi-axis.  The positions of the foci and
the condition that the ellipse passes through the point $w=(\mu,\Delta)$ are
two equations,
\begin{equation}\label{eq:ell2}
  \frac{S^{2}}4-h_E^{2}=E_p^{2},\qquad
  \frac{4\mu^{2}}{S^{2}}+\frac{\Delta^{2}}{h_E^{2}}=1 .
\end{equation}
The second does not contain the momentum.  Cleared of denominators it ties
the two semi-axes to each other,
\begin{equation}\label{eq:twoaxes}
  \Bigl(\frac{S^{2}}4-\mu^{2}\Bigr)\bigl(h_E^{2}-\Delta^{2}\bigr)=\mu^{2}\Delta^{2} ,
\end{equation}
and with $h_E$ taken from the first it is the quadratic \eqref{eq:quad} at
$z=S^{2}$.  As the momentum grows the foci move apart, the apex stays, and the
ellipse becomes longer and lower (Fig.~\ref{fig:twins}a).

\subsection{The twin}

\begin{figure}[t]
  \centering
  \includegraphics[width=0.96\linewidth]{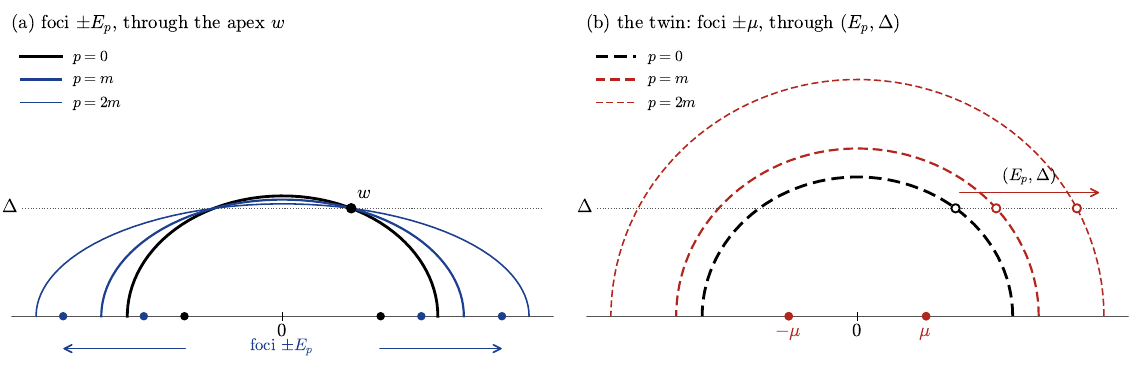}
  \caption{The two families of ellipses at the momenta $p=0$ (black), $m$ and
  $2m$, for the triangle of Fig.~\ref{fig:points}b.  (a)~The ellipse with foci
  $\pm E_p$ through the apex $w$: the foci move apart, the apex stays, and the
  minor semi-axis $h_E$ comes down towards $\Delta$.  (b)~Its twin, with foci
  $\pm\mu$ through the point $(E_p,\Delta)$: the foci stay, the point moves to
  the right, and the minor semi-axis $h_\mu$ grows.  At each momentum the two
  ellipses have the same major semi-axis, $\tfrac12S$.}
  \label{fig:twins}
\end{figure}

The levels $\sqrt{(E_p\pm\mu)^{2}+\Delta^{2}}$ do not change when $E_p$ and
$\mu$ are interchanged.  The same sum $S$ is therefore the major axis of a
second ellipse, the twin of the first,
\begin{equation}\label{eq:twin}
  \frac{S^{2}}4-h_\mu^{2}=\mu^{2},\qquad
  \frac{4E_p^{2}}{S^{2}}+\frac{\Delta^{2}}{h_\mu^{2}}=1 :
\end{equation}
its foci are $\pm\mu$, it passes through the point $(E_p,\Delta)$, and $h_\mu$
is its minor semi-axis.  The foci of the twin do not move.  As the momentum
grows the point moves to the right, and the twin grows
(Fig.~\ref{fig:twins}b).

\subsection{The circle and its two heights}

\begin{figure}[t]
  \centering
  \includegraphics[width=0.92\linewidth]{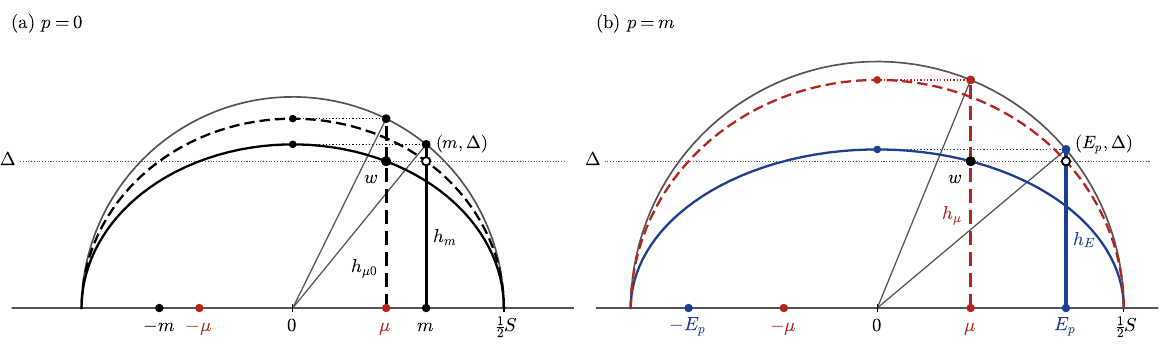}
  \caption{The circle and its two heights, for the triangle of
  Fig.~\ref{fig:points}b, (a)~at $p=0$ and (b)~at the momentum $p=m$.  The
  circle about the origin has the diameter $S=E^{+}+E^{-}$.  Above the point
  $E_p$ it stands at the height $h_E$, the minor semi-axis of the ellipse with
  foci $\pm E_p$ (solid), which passes through the apex $w$.  Above the point
  $\mu$ it stands at the height $h_\mu$, the minor semi-axis of the twin, with
  foci $\pm\mu$ (dashed), which passes through the point $(E_p,\Delta)$.
  Both points lie at the height $\Delta$ (dotted).  At $p=0$ the foci are
  $\pm m$ and $\pm\mu$, and the two heights are $h_m$ and $h_{\mu0}$ of
  Sec.~\ref{sec:arc}; as $p$ grows, $h_E$ comes down from $h_m$ towards
  $\Delta$ and $h_\mu$ grows from $h_{\mu0}$.  Thin lines: the radii from the
  origin to the tops of the two heights.}
  \label{fig:circle}
\end{figure}

We draw the circle about the origin whose diameter is $S$
(Fig.~\ref{fig:circle}).  Both ellipses lie inside it and touch it at the two
ends of the major axis, and above a focus of an ellipse the circle stands at a
height equal to the minor semi-axis of that ellipse: above the point $E_p$ it
stands at the height $h_E$, and above the point $\mu$, the foot of the height
of the triangle, at the height $h_\mu$.  Each height of the circle is named by
the point above which it stands.  The two heights obey one relation,
\begin{equation}\label{eq:heights}
  h_\mu^{2}=\frac{S^{2}}4-\mu^{2},\qquad
  h_E\,h_\mu=\tfrac12S\,\Delta .
\end{equation}
The second is \eqref{eq:twoaxes} again, and it follows from the figure.  An
ellipse is its circle compressed towards the major axis in the ratio of the
two semi-axes.  The point of the circle above $\mu$ stands at the height
$h_\mu$; compressed, it stands at the height $h_\mu h_E\big/\tfrac12S$, and
that point is the apex, at the height $\Delta$.  This is the law of the two
heights.

By \eqref{eq:quad}, \eqref{eq:ell2} and \eqref{eq:heights},
\begin{equation}\label{eq:prodheights}
  E^{+}E^{-}=h_E^{2}+h_\mu^{2}-\Delta^{2},\qquad
  h_E^{2}-h_\mu^{2}=\mu^{2}-E_p^{2} .
\end{equation}
The heights are equal where the points $E_p$ and $\mu$ coincide.  For $\mu>m$ that is the
Fermi sphere $E_p=\mu$: there the ellipse and its twin are one curve, and the
apex stands above its focus.  For $\mu<m$ the point $E_p$ never reaches $\mu$,
and $h_E<h_\mu$ at every momentum.

\subsection{What a mode holds}

Each momentum is a mode with two levels, and the circle is drawn for one mode.
Its radius and its two heights are the responses of the mode to the three
parameters: by Appendix~\ref{app:figure},
$\partial S/\partial\mu=4\mu\,h_E^{2}/(E^{+}E^{-}S)$,
$\partial S/\partial\Delta=\Delta\,S/(E^{+}E^{-})$, and
$dS/dp=4p\,h_\mu^{2}/(E^{+}E^{-}S)$, with $m$ in place of $p$ for the
derivative with respect to the mass.  These responses are what the mode
holds; we call them its contents.  At the
momentum $p$ the levels are the sides of the triangle with base $[-E_p,E_p]$
and apex $w$, and its angles at the base,
$\cos\theta_\pm(p)=(E_p\pm\mu)/E^{\pm}$, are the Bogoliubov angles of the two
branches.  We take them as full angles, as in Anderson's picture of
pseudospins~\cite{Anderson1958}, so that $u_\pm=\cos\tfrac12\theta_\pm$ and
$v_\pm=\sin\tfrac12\theta_\pm$: $v_-^{2}$ is the occupation of the particle
state of the mode, $v_+^{2}$ that of the antiparticle state, and $u_\pm v_\pm$
are the two pair amplitudes.  Half the derivatives of $S$ with respect to
$\mu$, $E_p$ and $\Delta$ are then
\begin{equation}\label{eq:modes}
  v_-^{2}-v_+^{2}=\frac{2\,\mu\,h_E^{2}}{E^{+}E^{-}S},\qquad
  1-v_-^{2}-v_+^{2}=\frac{2\,E_p\,h_\mu^{2}}{E^{+}E^{-}S},\qquad
  u_-v_-+u_+v_+=\frac{2\,\Delta\,(S/2)^{2}}{E^{+}E^{-}S} .
\end{equation}
The first is the charge of the mode, particles minus antiparticles.  The
second, times $m/E_p$, is its scalar density: in the vacuum it is the term of
the Dirac sea, and each particle or antiparticle present blocks part of it.
The third is its pair amplitude.  The square of the height above $E_p$ thus
goes with the charge, the square of the height above $\mu$ with the scalar
density, and the square of the radius with the pairs.

\section{The integral on an arc}\label{sec:arc}

As the momentum grows from zero to infinity the foci of the ellipse move
apart and the circle grows.  One of its two heights grows with it without
bound, and the other stays between two fixed values.

\subsection{The path}

At $p=0$ the foci are the ends of the base and the diameter is
$\ell_++\ell_-$.  The two heights are then $h_m$ above the point $m$ and
$h_{\mu0}$ above the point $\mu$ (Fig.~\ref{fig:circle}a), with
$h_m^{2}=\tfrac14(\ell_++\ell_-)^{2}-m^{2}$ and
$h_{\mu0}^{2}=\tfrac14(\ell_++\ell_-)^{2}-\mu^{2}$.  The point $\mu$ stays
where it is, and the height above it grows with the circle.  With
$\rho^{2}\equiv m^{2}+\Delta^{2}$, eliminating $S$ and $h_E$ between the
equations of the ellipse \eqref{eq:ell2} and the law \eqref{eq:heights} gives
\begin{equation}\label{eq:p2h}
  p^{2}=h_\mu^{2}+\mu^{2}-\rho^{2}-\frac{\mu^{2}\Delta^{2}}{h_\mu^{2}} ,
\end{equation}
so that at large momenta $h_\mu$ is the momentum itself.  The height above
$E_p$ decreases instead, from $h_m$ towards $\Delta$, and in terms of it the
momentum line is the finite interval $\Delta<h_E\le h_m$.  On this interval
$p^{2}$ is a ratio, \eqref{eq:p2beta}, whose numerator has two factors,
$h_m^{2}-h_E^{2}$ and $h_E^{2}+h_{\mu0}^{2}-\Delta^{2}$.  The first suggests a
cosine, and the second is then Legendre's square root:
\begin{equation}\label{eq:betaphi}
  \begin{gathered}
  h_E=h_m\cos\varphi:\qquad
  \sqrt{h_m^{2}-h_E^{2}}=h_m\sin\varphi,\\
  \sqrt{h_E^{2}+h_{\mu0}^{2}-\Delta^{2}}=\sqrt{\ell_+\ell_-}\,\sqrt{1-k^{2}\sin^{2}\varphi},\qquad
  k^{2}=\frac{h_m^{2}}{\ell_+\ell_-} .
  \end{gathered}
\end{equation}
The modulus $k$ belongs to the triangle of Fig.~\ref{fig:points}b,
\begin{equation}\label{eq:kdict}
  k^{2}=\frac{(\ell_++\ell_-)^{2}-4m^{2}}{4\,\ell_+\ell_-}=\cos^{2}\frac\theta2 ,
\end{equation}
the cosine of half the angle at the apex.  Legendre's amplitude $\varphi$ has a
meaning in the figure as well: $\cos\varphi$ is the height $h_E$ in units of its largest
value.  It runs from $\varphi=0$ at $p=0$ to $\varphi_1$ at $p=\infty$, where
$h_E=\Delta$:
\begin{equation}\label{eq:phi1}
  \cos\varphi_1=\frac{\Delta}{h_m},\qquad
  \sin\varphi_1=\frac{2\mu}{\ell_++\ell_-} .
\end{equation}

\begin{figure}[t]
  \centering
  \includegraphics[width=0.94\linewidth]{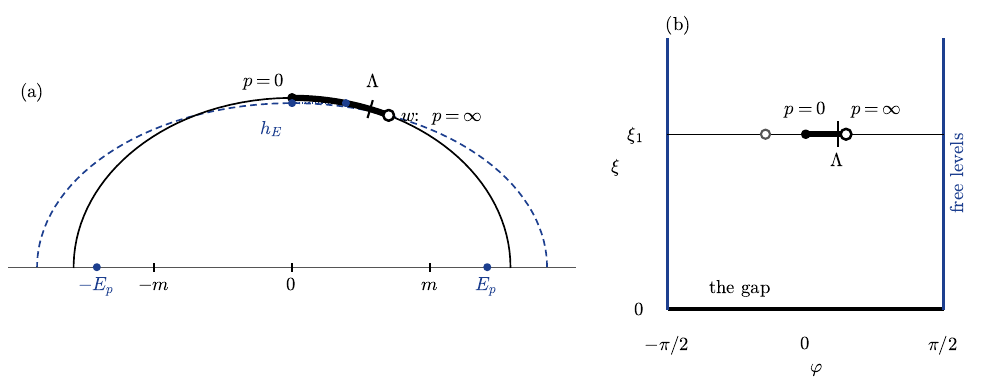}
  \caption{The momentum line as an arc, for the triangle of
  Fig.~\ref{fig:points}b.  (a)~The ellipse of $p=0$ (solid), with foci
  $\pm m$, and the ellipse of the momentum $p=m$ (dashed), with foci
  $\pm E_p$; both pass through the apex $w$.  The dotted line carries the
  height $h_E$ of the top of the dashed ellipse to the solid one.  The thick
  arc is the set of such marks for all momenta; it ends at the apex, and the
  bar shows where the upper limit $\Lambda=2m$ stops it.  (b)~The same plane in
  the coordinates $x+iy=m\sin(\varphi+i\xi)$.  The gap of the free spectrum is
  the floor, the free levels $\pm E_p$ are the walls, and the arc is a
  straight segment.  The grey circle is the mirror image $(-\mu,\Delta)$ of
  the apex.}
  \label{fig:arc}
\end{figure}

Figure~\ref{fig:arc} draws the path.  The ellipse of $p=0$, with foci
$\pm m$ and semi-axes $\tfrac12(\ell_++\ell_-)$ and $h_m$, does not depend on
the momentum.  On it we mark the point that stands at the height $h_E$, the
height of the point of the circle above $E_p$.  As $p$ grows, the mark
descends from the top of the ellipse to the apex, which it reaches at
$p=\infty$.  The whole half-line $p\ge0$ is this arc
(Fig.~\ref{fig:arc}a).  In the elliptic coordinates of the plane,
$x+iy=m\sin(\varphi+i\xi)$, the ellipse is the line $\xi=\xi_1$ with
$\cosh\xi_1=(\ell_++\ell_-)/(2m)$, and the arc is a straight segment,
$0\le\varphi\le\varphi_1$ (Fig.~\ref{fig:arc}b).
The ellipse through the apex at $p=0$ thus carries the path, while the twin,
whose foci do not move, carries the height $h_\mu$ that grows with the momentum.  A finite upper limit
stops the mark at an amplitude $\varphi_\Lambda$ short of the apex, at the
distance
\begin{equation}\label{eq:sLam}
  \sin^{2}\varphi_1-\sin^{2}\varphi_\Lambda
  =\frac{\mu^{2}\cos^{2}\varphi_1}{h_\mu^{2}(\Lambda)}
  =\frac{\mu^{2}\cos^{2}\varphi_1}{\Lambda^{2}}+O(\Lambda^{-4}) ,
\end{equation}
so that large momenta are the neighbourhood of the end of the arc.

\subsection{The step and three weights}

\begin{figure}[t]
  \centering
  \includegraphics[width=0.90\linewidth]{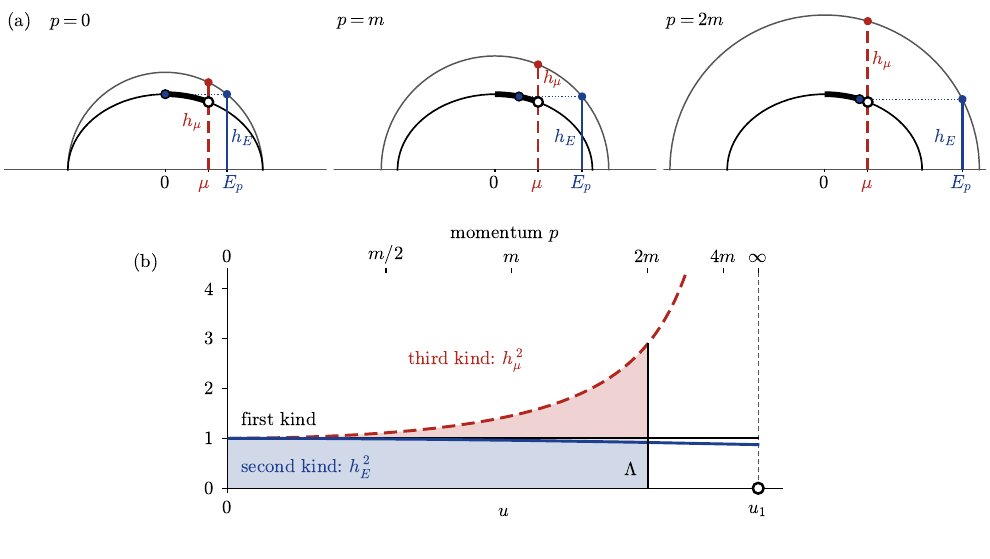}
  \caption{The three weights, for the triangle of
  Fig.~\ref{fig:points}b.  (a)~Three momenta.  The ellipse of $p=0$ and the
  arc of Fig.~\ref{fig:arc} do not move; the circle about the origin, of
  diameter $S$, grows.  Its height above the point $E_p$ is $h_E$, the height of the
  mark on the arc (dotted line); its height above the point $\mu$ is $h_\mu$.
  The first descends to the height $\Delta$ of the apex (open circle), the
  second grows without bound.  (b)~The three weights of \eqref{eq:weights}
  along the path, in the coordinate $u$ in which the step is $du$; the upper
  axis is the momentum.  $\EllF$, $\EllE$ and $\Pi$ are the areas under the
  line, under the lower curve and under the upper curve, shaded up to
  $\Lambda=2m$.  The path ends at $u_1$, where the upper curve has a simple
  pole.}
  \label{fig:weights}
\end{figure}

Along the arc the amplitude advances with the momentum, by
\eqref{eq:onpath} and \eqref{eq:prodheights}, as
\begin{equation}\label{eq:step}
  \frac{d\varphi}{\sqrt{1-k^{2}\sin^{2}\varphi}}=2\mu\sqrt{\ell_+\ell_-}\;\frac{dp}{E^{+}E^{-}S},
  \qquad\text{so that}\qquad
  u\equiv\EllF(\varphi,k)=2\mu\sqrt{\ell_+\ell_-}\int_0^{p}\frac{dp'}{E^{+}E^{-}S} .
\end{equation}
Legendre's integral of the first kind is thus the momentum integrated with the
factor $1/(E^{+}E^{-}S)$.  At large $p$ this factor falls as $1/(2p^{3})$,
which is why the whole half-line of the momentum has a finite length in this
coordinate, $u_1=\EllF(\varphi_1,k)$.  In the coordinate $u$ the three integrals of
Legendre differ by a weight and by nothing else:
\begin{equation}\label{eq:weights}
  \begin{gathered}
  \EllF(\varphi,k)=\int_0^{u}du',\qquad
  \EllE(\varphi,k)=\int_0^{u}\frac{h_E^{2}+h_{\mu0}^{2}-\Delta^{2}}{\ell_+\ell_-}\,du',\\
  \Pi(\varphi;\alpha^{2},k)=\int_0^{u}\frac{h_\mu^{2}}{h_{\mu0}^{2}}\,du',
  \qquad \alpha^{2}=\frac1{\sin^{2}\varphi_1} ,
  \end{gathered}
\end{equation}
where $\Pi(\varphi;\alpha^{2},k)=\int_0^{\varphi}d\vartheta\,
(1-\alpha^{2}\sin^{2}\vartheta)^{-1}(1-k^{2}\sin^{2}\vartheta)^{-1/2}$.  The
weights are unity, the square of the height of the mark plus a constant, and
the square of the height above $\mu$; the last two are normalized to unity at
$p=0$ (Fig.~\ref{fig:weights}).  The first two integrals are finite because the
path has a finite length.  The third grows without bound: $h_\mu^{2}$ grows as
$p^{2}$, the step falls as $dp/(2p^{3})$, and their product is $dp/(2p)$, a
logarithm.

\subsection{The integral at any upper limit}

One integration by parts removes the factor $p^{2}$.  We look for a boundary
term that removes everything in $p^{2}S$ that grows and leaves a combination
of the three weights times the step $dp/(E^{+}E^{-}S)$.
Appendix~\ref{app:path} gives it:
\begin{equation}\label{eq:oneline}
  p^{2}S
  =\frac{d}{dp}\Bigl[\frac{p\,S\,(16\,p^{2}+10\,\rho^{2}-S^{2})}{48}\Bigr]
  +\frac{c_1+c_2\,h_E^{2}+c_3\,h_\mu^{2}}{E^{+}E^{-}\,S} ,
\end{equation}
with coefficients $c_1$, $c_2$ and $c_3$ that are polynomials in $m$, $\mu$
and $\Delta$, \eqref{eq:coefs}.  This is an identity between
algebraic functions of the momentum, and checking it takes one
differentiation.  At $\mu=0$ it is the integration by parts of the free
vacuum energy.  Reference~\cite{companion} arrives at it by another route,
Hermite's reduction of the integral in the variable $S^{2}$.  Integrated along
the arc with the step \eqref{eq:step} and the weights \eqref{eq:weights}, it
gives
\begin{equation}\label{eq:closed}
  \begin{split}
  I(\Lambda)=W(\Lambda)+\frac{1}{2\mu\sqrt{\ell_+\ell_-}}\Bigl[&
  \bigl(c_1-c_2\,(h_{\mu0}^{2}-\Delta^{2})\bigr)\,\EllF(\varphi_\Lambda,k)
  +c_2\,\ell_+\ell_-\,\EllE(\varphi_\Lambda,k)\\
  &+c_3\,h_{\mu0}^{2}\,\Pi(\varphi_\Lambda;\alpha^{2},k)\Bigr],
  \qquad \cos\varphi_\Lambda=\frac{h_E(\Lambda)}{h_m} ,
  \end{split}
\end{equation}
exact at every $\Lambda$.  Here $W$ is the boundary term of
\eqref{eq:oneline}, and $\varphi_\Lambda$ is the amplitude at which the upper
limit stops the mark.

Let $T_1$, $T_2$ and $T_3$ be the sums of $1$, $h_E^{2}$ and $h_\mu^{2}$ over
the momentum with the step $dp/(E^{+}E^{-}S)$; they are combinations of
$\EllF$, $\EllE$ and $\Pi$, \eqref{eq:Tsums}.  By \eqref{eq:modes}, the three
contents of a mode summed with the measure $dp$ of one space dimension are
\begin{equation}\label{eq:holds}
  \begin{alignedat}{2}
  &\text{charge:}\quad&&\int_0^{\Lambda}\bigl(v_-^{2}-v_+^{2}\bigr)\,dp=2\mu\,T_2,\\[6pt]
  &\text{scalar density:}\quad&&\int_0^{\Lambda}\frac{m}{E_p}\bigl(1-v_-^{2}-v_+^{2}\bigr)\,dp=2m\,T_3,\\[6pt]
  &\text{pairs:}\quad&&\int_0^{\Lambda}\bigl(u_-v_-+u_+v_+\bigr)\,dp=2\Delta\bigl(\mu^{2}T_1+T_3\bigr).
  \end{alignedat}
\end{equation}
Here $T_1$ is the length of the path in the coordinate $u$, $T_2$ gives the
charge and $T_3$ the scalar density; the pairs go with the radius and combine
$T_1$ and $T_3$.  In one dimension these are, up to
the factors of the potential, the density, the scalar condensate and the gap
equation.  With the measure $p^{2}dp$ of three dimensions the same three sums
and the boundary term make up the integral itself, \eqref{eq:physclosed}.  At
large momenta the charge of a mode falls as $\mu\Delta^{2}/p^{3}$, while its
scalar density keeps the value $m/E_p$ of the vacuum: the sum that grows is
that of the scalar density.

\subsection{Three growing terms}

Everything in \eqref{eq:closed} that grows with $\Lambda$ comes from the end
of the arc.  The boundary term contains two powers,
\begin{equation}\label{eq:WLam}
  W(\Lambda)=\frac{\Lambda^{4}}{2}+\frac{\rho^{2}\Lambda^{2}}{2}+\mathcal C_W+O(\Lambda^{-2}),
  \qquad
  \mathcal C_W=\frac{\mu^{2}\Delta^{2}}{12}+\frac{\rho^{4}}{16} ,
\end{equation}
and the chemical potential drops out of both.  It enters what grows only
through the coefficient of the logarithm, which is fixed by the integrand at
large momenta,
\begin{equation}\label{eq:Sasym}
  S=2p+\frac{\rho^{2}}{p}+\frac{c_3/2}{p^{3}}+O(p^{-5}),\qquad
  \frac{c_3}{2}=\mu^{2}\Delta^{2}-\frac{\rho^{4}}{4} .
\end{equation}
The logarithm itself is in the sum of the third weight, where the pole of
$\Pi$ at the end of the arc can be exchanged for an elementary function
(Appendix~\ref{app:end}).  The result is
\begin{equation}\label{eq:asymp}
  I(\Lambda)=\frac{\Lambda^{4}}{2}+\frac{\rho^{2}\Lambda^{2}}{2}
  +\frac{c_3}{2}\ln\frac{2\Lambda}{\ell_0}+\mathcal C(m,\mu,\Delta)+O(\Lambda^{-2}),\qquad
  \ell_0\equiv\sqrt{\ell_+\ell_-+\mu^{2}} ,
\end{equation}
with the finite part
\begin{equation}\label{eq:Cfin}
  \mathcal C(m,\mu,\Delta)=\mathcal C_W+\frac{1}{2\mu\sqrt{\ell_+\ell_-}}\Bigl[
  \bigl(c_1-c_2\,(h_{\mu0}^{2}-\Delta^{2})+c_3\,h_{\mu0}^{2}\bigr)\,\EllF(\varphi_1,k)
  +c_2\,\ell_+\ell_-\,\EllE(\varphi_1,k)
  -c_3\,h_{\mu0}^{2}\,\Pi(\varphi_1;\omega^{2},k)\Bigr] ,
\end{equation}
where $\omega^{2}=k^{2}\sin^{2}\varphi_1$.  As the upper limit grows, the
integral thus grows by two powers and one logarithm, and the coefficient of
the logarithm is a polynomial in the parameters.  The remainder, Legendre's
three integrals taken over the whole arc, is finite.

\subsection{The integral in closed form}

Where the three growing terms go is decided by the potential in which the
integral appears.  It stands there together with a term quadratic in the gap,
$\Delta^{2}$ divided by the coupling, and at a fixed mass the three terms
regroup accordingly,
\begin{equation}\label{eq:split}
  \frac{\Lambda^{4}}{2}+\frac{\rho^{2}\Lambda^{2}}{2}+\frac{c_3}{2}\ln\Lambda
  =\Bigl[\frac{\Lambda^{4}}{2}+\frac{m^{2}\Lambda^{2}}{2}-\frac{m^{4}}{4}\ln\Lambda\Bigr]
  +\Delta^{2}\Bigl[\frac{\Lambda^{2}}{2}+\frac{m^{2}}{2}\ln\Lambda\Bigr]
  +\Bigl[\Delta^{2}\bigl(\mu^{2}-m^{2}\bigr)-\frac{\Delta^{4}}{4}\Bigr]\ln\Lambda .
\end{equation}
The first bracket is the growing part of the zero-point energy of free
fermions of mass $m$ and does not depend on the state.  The second is one
function of $\Lambda$ times $\Delta^{2}$, which is what a coupling absorbs.
The third cannot be absorbed in this way: a coupling does not depend on the
state, and the terms with $\mu^{2}\Delta^{2}$ and $\Delta^{4}$ are not one
function of $\Lambda$ times $\Delta^{2}$.  How the term
$-m^{2}\Delta^{2}\ln\Lambda$ is divided between the second bracket and the
third is a convention, here that of the threshold of two particles
(Sec.~\ref{sec:crossover}).  The two powers thus go into the constant and into
the coupling, each with a part of the logarithm, and a logarithm stays.  When
the mass is itself a condensate, as in the Nambu--Jona-Lasinio model
(Sec.~\ref{sec:pion}), the coupling multiplies $\rho^{2}$: it takes
$\rho^{2}\Lambda^{2}$ whole, and no part of the logarithm is absorbed.

The way the integral is stopped enters the finite part through one number,
the scale of the logarithm.  A rule that
acts on the momentum, a window $f(p/\Lambda)$ that is one at small momenta and
falls off at large ones, touches the integral near the end of the arc only,
and
\begin{equation}\label{eq:anyrule}
  \int_0^{\infty}p^{2}S\,f(p/\Lambda)\,dp
  =2f_3\,\Lambda^{4}+f_1\,\rho^{2}\Lambda^{2}
  +\frac{c_3}{2}\ln\frac{2s_f\Lambda}{\ell_0}+\mathcal C+o(1),
\end{equation}
with three numbers $f_3$, $f_1$ and $s_f$ that depend on the window
alone~\cite{supp}; for the upper limit of \eqref{eq:asymp} they are
$\tfrac14$, $\tfrac12$ and $1$.
The rule changes the coefficients of the two powers, which multiply $1$ and
$\rho^{2}$ whatever it is, and it moves the scale of the logarithm by the
factor $s_f$.  It changes nothing else: the coefficient of the logarithm is
the same, and for the finite part every such rule acts as the upper limit
$s_f\Lambda$.  The rule can therefore be replaced by its scale: we integrate
over all momenta,
subtract the two powers and measure the logarithm from a scale $\lambda$,
\begin{equation}\label{eq:Ilam}
  I_\lambda(m,\mu,\Delta)\equiv\lim_{\Lambda\to\infty}\Bigl[I(\Lambda)-\frac{\Lambda^{4}}{2}
  -\frac{\rho^{2}\Lambda^{2}}{2}-\frac{c_3}{2}\ln\frac{\Lambda}{\lambda}\Bigr] .
\end{equation}
By \eqref{eq:asymp} what is left of the integral is
\begin{equation}\label{eq:main}
  I_\lambda(m,\mu,\Delta)=\mathcal C(m,\mu,\Delta)+\frac{c_3}{2}\ln\frac{2\lambda}{\ell_0} .
\end{equation}
Equation~\eqref{eq:main} is the main result of this paper.  The finite part
\eqref{eq:Cfin} holds Legendre's three integrals at the end of the arc, the
coefficient of the logarithm is the polynomial of \eqref{eq:Sasym}, and
$\lambda$ is the one number that the integral does not fix.  From here on,
``with all momenta integrated'' means this limit, for the integral and for
everything derived from it: the powers are subtracted and $\ln\Lambda$ is
replaced by $\ln\lambda$.  Up to terms of the order $\lambda^{-2}$ it is the
integral stopped at $\Lambda=\lambda$, without its two powers.

The energy $\ell_0=\sqrt{\ell_+\ell_-+\mu^{2}}$ under the logarithm is not a
parameter.  It is fixed by the triangle of Fig.~\ref{fig:points}b, and in the
vacuum, at $\Delta=0$ and $\mu<m$, it is the mass, so that the logarithm is
$\ln(2\lambda/m)$.  This is where the effective range $r_e$ of the channel of
the pair will be read, $\ln(2\lambda/m)=1-\tfrac14\pi\,m\,r_e$
(Sec.~\ref{sec:range}).  How \eqref{eq:main} is divided between the finite
part and the logarithm is a convention: with another energy $\ell_0'$ in place
of $\ell_0$ the finite part becomes $\mathcal C+\tfrac12c_3\ln(\ell_0'/\ell_0)$,
the sum is the same, and the point $2\lambda=\ell_0$ has no meaning of its
own.  What does not depend on the convention is the coefficient $\tfrac12c_3$
and the fact that the integral depends on $\lambda$ through one logarithm,
$\lambda\,\partial I_\lambda/\partial\lambda=\tfrac12c_3$.  A change of the
scale itself is not a convention once the integral enters a potential: the
terms $\mu^{2}\Delta^{2}-\tfrac14\Delta^{4}$ of $\tfrac12c_3$ are neither
constant nor proportional to $\Delta^{2}$ alone, and neither the constant nor
the coupling can undo them.  The gap equation and the density of the crossover
are derivatives of \eqref{eq:main} (Sec.~\ref{sec:crossover}).

\section{The crossover in closed form}\label{sec:crossover}

The potential of a mean-field theory gives its equations by differentiation:
the derivative with respect to the gap is the gap equation, and the derivative
with respect to the chemical potential is the density.  As in Sec.~\ref{sec:arc}, we keep the
upper limit $\Lambda$ until the growing terms are in hand and then integrate
over all momenta.

\subsection{The two equations}

Half the derivatives of the integral \eqref{eq:integral} in $\mu$, $m$ and
$\Delta$ are the three contents \eqref{eq:modes} of a mode, summed over the
modes with the factor $p^{2}$:
\begin{equation}\label{eq:Qdef}
  Q_\mu\equiv\int_0^{\Lambda}p^{2}\bigl(v_-^{2}-v_+^{2}\bigr)dp,\qquad
  Q_m\equiv\int_0^{\Lambda}p^{2}\,\frac{m}{E_p}\bigl(1-v_-^{2}-v_+^{2}\bigr)dp,\qquad
  Q_\Delta\equiv\int_0^{\Lambda}p^{2}\bigl(u_-v_-+u_+v_+\bigr)dp .
\end{equation}
Up to the factor that multiplies the integral in a potential, $Q_\mu$ is the
density of the charge, $Q_m$ the scalar density, and $Q_\Delta$ the integral
of the gap equation.  The factor $p^{2}$ is taken away as in
Sec.~\ref{sec:arc}, and each of the three is then one boundary term and a
combination of the same three sums along the arc, \eqref{eq:Qclosed}.  We
write the potential in which the integral stands as
$\Omega=\Delta^{2}/G-N\,I(\Lambda)$, where the coupling $G$ and the number $N$
depend on the model.  The gap equation
$\partial\Omega/\partial\Delta=0$ is then $Q_\Delta/\Delta=1/(NG)$, and since
$u_\pm v_\pm=\Delta/2E^{\pm}$, its left-hand side is
$\tfrac12\int_0^{\Lambda}p^{2}(1/E^{+}+1/E^{-})\,dp$.  At a given density the
gap equation and the density, $Q_\mu$ times the factor of the model, are two
equations for $\mu$ and $\Delta$.  They are the equations of the mean-field
relativistic crossover~\cite{HeZhuang2007,Brauner2008}.

\subsection{The coupling and the mass of the pair}

The coupling can be traded for a property of the pair.  As $\Delta\to0$ at
$\mu<m$ the two levels become $E_p\pm\mu$, and the left-hand side of the gap
equation becomes the integral of $E_p\,p^{2}/(p^{2}+q^{2})$ with
$q^{2}=m^{2}-\mu^{2}$, the loop of two free particles at the total energy
$2\mu$.  With $1/(NG)$ subtracted, this loop is the denominator of the
scattering amplitude of the two particles in the vacuum
(Sec.~\ref{sec:range}), and a bound pair is its zero.  For a bound pair, let
$m_B$ be the value
of $2\mu$ at which the gap sets in.  It is the mass of the pair, which
condenses when its chemical potential reaches its
mass~\cite{NishidaAbuki2005}.  The gap equation at the onset is the
condition for the pole of the pair, the Thouless
criterion~\cite{Thouless1960},
\begin{equation}\label{eq:Vdef}
  \begin{gathered}
  \frac{1}{NG}=V(\Lambda)\equiv\int_0^{\Lambda}\frac{E_p\,p^{2}}{p^{2}+q_B^{2}}\,dp,\qquad
  q_B^{2}\equiv m^{2}-\tfrac14m_B^{2},\\
  V(\Lambda)=\frac{\Lambda E_\Lambda+m^{2}\operatorname{arsinh}(\Lambda/m)}{2}
  -q_B^{2}\operatorname{arsinh}\frac{\Lambda}{m}
  -q_B\sqrt{m^{2}-q_B^{2}}\,\Bigl[\frac{\pi}{2}-\arctan\frac{q_B\,E_\Lambda}{\Lambda\sqrt{m^{2}-q_B^{2}}}\Bigr] ,
  \end{gathered}
\end{equation}
where $q_B>0$ is the binding momentum of the pair and
$E_\Lambda=\sqrt{\Lambda^{2}+m^{2}}$.  With the coupling eliminated, the gap
equation at the upper limit $\Lambda$ is the zero of the denominator
\begin{equation}\label{eq:gapren}
  D(\Lambda)\equiv\frac{Q_\Delta}{\Delta}-V(\Lambda)=0 .
\end{equation}
Up to a sign, $D$ is the inverse propagator of the pair at zero
momentum~\cite{Abuki2007}.

The coupling can equally be traded at the threshold of two particles,
$q_B=0$, where $V$ is the integral of $E_p$, whose growing part multiplies
$\Delta^{2}$ in \eqref{eq:split}.  This is the trade for a scattering
length~\cite{Abuki2007,HeZhuang2007,Brauner2008}.  With the scattering length
$a$ defined by $(\pi m/2)/a\equiv V|_{q_B=0}-1/(NG)$, the gap equation is
\begin{equation*}
  D\big|_{q_B=0}+\frac{\pi m}{2a}=0 .
\end{equation*}
Where the pair is bound this is \eqref{eq:gapren}, and it holds for either
sign of $a$.

\begin{figure}[t]
  \centering
  \includegraphics[width=0.74\linewidth]{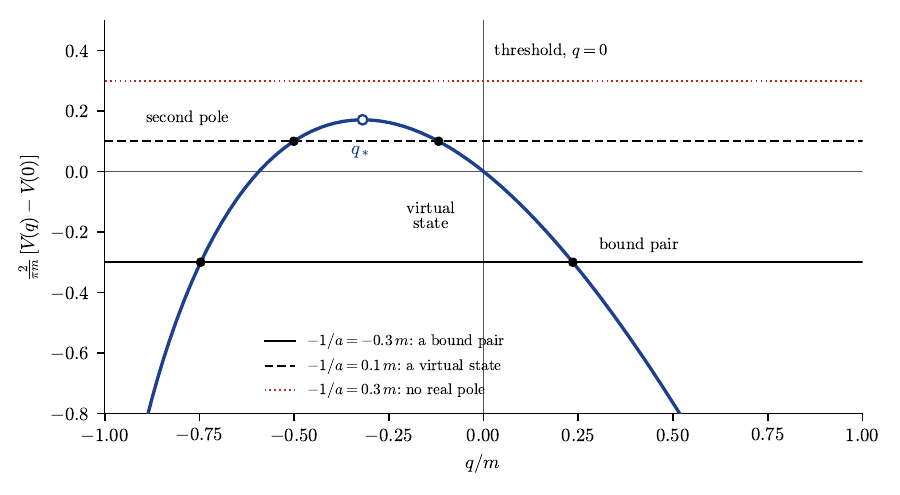}
  \caption{The poles of the pair on the curve of the loop:
  $\tfrac{2}{\pi m}[V(q)-V(0)]$ against $q$ at $\Lambda=10\,m$, where $V(q)$ is
  the closed form of \eqref{eq:Vdef} at $q_B=q$.  The coupling is the
  horizontal line at $\tfrac{2}{\pi m}[1/(NG)-V(0)]=-1/a$, and the poles are
  its crossings with the curve: a bound pair and a far pole (solid line), a
  virtual state and a second pole (dashed), and no real pole above the maximum
  of the curve at $q_*$ (dotted).}
  \label{fig:pole}
\end{figure}

A weaker attraction binds nothing.  The pole then passes the threshold,
$q_B=0$, and goes on to the second sheet of the energy: a virtual state at
$q_B<0$, as for a negative scattering length.  The closed form of
\eqref{eq:Vdef}, continued through $q_B=0$, gives the coupling of this weaker
attraction, and $m_B$ is then the energy of the virtual state and not a mass.
The virtual state stays on the real axis as long as $V$ grows with $-q_B$
(Fig.~\ref{fig:pole}).  At the maximum of $V$ on this side, $q_B=q_*$, it
meets a second pole, and the two leave the axis together; a still weaker
attraction is not described by a real $q_B$, and the gap equation is then used
in the form with $a$.  Section~\ref{sec:range} relates $q_*$ to the effective
range of the channel.

\subsection{The logarithm and the scale}

With the coupling traded for the pair, the two equations are $D=0$ and the
density.  In the nonrelativistic crossover the same trade leaves both
equations finite when all momenta are integrated, and the scattering length is
then the only parameter of the
attraction~\cite{MariniPistolesiStrinati1998}.  Here each of the two keeps one
logarithm.  The integrands of the two equations are
$f_D=\tfrac12p^{2}(1/E^{+}+1/E^{-})-E_p\,p^{2}/(p^{2}+q_B^{2})$ and
$f_\mu=p^{2}(v_-^{2}-v_+^{2})$.  At momenta large compared with $|q_B|$ and
$\sqrt{|m^{2}-\mu^{2}|}$, where the gap is small compared with the levels, the
leading term of $f_D$ is $(\mu^{2}-m^{2}+q_B^{2})\,E_p/p^{2}$ and $f_\mu$ is
$\mu\Delta^{2}\,E_p/p^{2}$: one factor, $E_p/p^{2}$, multiplies both.  Below the mass this factor is
$m/p^{2}$: these are the tails of the nonrelativistic problem, and they
converge.  Above the mass it is $1/p$, and with the term $-\Delta^{2}/2p$ that
the gap adds to $f_D$,
\begin{equation}\label{eq:tails}
  f_D\;\to\;\frac1p\Bigl(\mu^{2}-\tfrac14m_B^{2}-\tfrac12\Delta^{2}\Bigr),\qquad
  f_\mu\;\to\;\frac1p\,\mu\Delta^{2}\qquad(p\gg m).
\end{equation}
A tail $1/p$ integrates to a logarithm.  The difference from the
nonrelativistic problem is the growth of the energy with the momentum.  The
nonrelativistic energy grows as $p^{2}$, so each inverse power of the energy
suppresses the integrand by two powers of the momentum.  The relativistic
energy grows as $p$, so each inverse power suppresses it by only one.

With all momenta integrated the two equations follow from \eqref{eq:main}
(Appendix~\ref{app:constants}), and the logarithm of each is measured from the
scale $\lambda$,
\begin{equation}\label{eq:Qinf}
  Q_\mu=\mu\Delta^{2}\ln\frac{2\lambda}{\ell_0}+q_\mu,\qquad
  D=\bigl(\mu^{2}-\tfrac14m_B^{2}-\tfrac12\Delta^{2}\bigr)\ln\frac{2\lambda}{\ell_0}+D_0 ,
\end{equation}
with $q_\mu$ and $D_0$ of \eqref{eq:qmuD0}, which do not depend on the scale
and hold Legendre's integrals at the end of the arc.  These are \eqref{eq:Qdef} and
\eqref{eq:gapren} with all momenta integrated: the powers cancel between
$Q_\Delta/\Delta$ and $V$, $Q_\mu$ has none, and each keeps its logarithm.
The coefficients of the two logarithms are those of the tails
\eqref{eq:tails}, and they are derivatives of one function,
\begin{equation}\label{eq:bosegas}
  \begin{gathered}
  \mu^{2}-\tfrac14m_B^{2}-\tfrac12\Delta^{2}=-\frac{\partial\Omega_B}{\partial(\Delta^{2})},\qquad
  \mu\Delta^{2}=-\frac12\frac{\partial\Omega_B}{\partial\mu},\\
  \Omega_B\equiv\Delta^{2}\bigl(\tfrac14m_B^{2}-\mu^{2}\bigr)+\tfrac14\Delta^{4}
  =\tfrac14\bigl[(m_B^{2}-\mu_B^{2})\,\Delta^{2}+\Delta^{4}\bigr],\qquad \mu_B=2\mu .
  \end{gathered}
\end{equation}
At any gap, $\Omega_B$ is the mean-field potential of a relativistic Bose
gas~\cite{Kapusta1981}: a field with the chemical potential $\mu_B=2\mu$, the
mass $m_B$ and a quartic repulsion.  For $q_B>0$ it is the bound pair itself,
which condenses when $\mu_B$ exceeds $m_B$.  Its kinetic term and its
quartic term come with one common logarithm, the signature of a composite
field~\cite{Eguchi1976,BardeenHillLindner1990}, and the scale is left only in
the normalization of this field.  In the Nambu--Jona-Lasinio model, near the
onset of the condensation, the whole potential reduces to that of a Bose gas
of pairs~\cite{He2010}; here $\Omega_B$ is what the logarithm of the scale
multiplies at any gap (Appendix~\ref{app:constants}).

The scale is the second parameter of the attraction, next to the scattering
length, and so the third parameter of the crossover.  Section~\ref{sec:range}
shows that it fixes the effective range of the channel of the pair.  At a given
density, \eqref{eq:Qinf} are two
equations for $\mu$ and $\Delta$ in closed form: a solution is a common root
of two closed expressions, and no momentum integral has to be computed.  As in
Ref.~\cite{companion}, the state is the local minimum that continues from
small gaps: with the logarithm measured from $\lambda$ the potential turns down
at gaps of the order of $\lambda$ and is not bounded below.

\section{The cutoff as an effective range}\label{sec:range}

In the literature on the relativistic crossover the coupling is traded for a
scattering length, as in the nonrelativistic problem, and the logarithm that is
left is cut off, so that the cutoff stays in the equations as a
parameter~\cite{Brauner2008,HeZhuang2007}.  In the language of
Sec.~\ref{sec:crossover} this parameter is the scale $\lambda$.  We show here
that it fixes the effective range of the channel of the pair, and that at low
density the crossover is that of a Feshbach resonance with this range.

\subsection{The amplitude of two particles}

The loop $V$ of \eqref{eq:Vdef} does more than fix the coupling.  With $q_B$
replaced by a variable $q$ it is the vacuum loop of two particles at the total
energy $2\sqrt{m^{2}-q^{2}}$, which we write $V(q)$.  Since $1/(NG)=V(q_B)$,
the difference $V(q_B)-V(q)$ is the denominator of their scattering amplitude.  We take both
loops with all momenta integrated (Appendix~\ref{app:constants}),
\begin{equation*}
  V_\lambda(q)=\frac{m^{2}}{4}+\bigl(\tfrac12m^{2}-q^{2}\bigr)\ln\frac{2\lambda}{m}
  -q\sqrt{m^{2}-q^{2}}\,\arccos\frac{q}{m} ,
\end{equation*}
so that the scattering length of Sec.~\ref{sec:crossover}, defined at the
threshold of two particles as in Ref.~\cite{HeZhuang2007}, is
$(\pi m/2)/a=V_\lambda(0)-V_\lambda(q_B)$.  Near the threshold
\begin{equation}\label{eq:Dexp}
  \frac{2}{\pi m}\bigl[V_\lambda(q_B)-V_\lambda(q)\bigr]=-\frac{1}{a}+q-\tfrac12\,r_e\,q^{2}+O(q^{3}),\qquad
  r_e=-\frac{4}{\pi m}\Bigl[\ln\frac{2\lambda}{m}-1\Bigr] .
\end{equation}
A bound pair is the zero at $q=q_B$.  Above the threshold, at $q=-ik$, the
left-hand side,
multiplied by the factor $m/\sqrt{m^{2}+k^{2}}$ of the relativistic phase
space, is $k\cot\delta-ik$, where $\delta$ is the phase shift of two particles of
relative momentum $k$, so that
\begin{equation}\label{eq:kcot}
  k\cot\delta=-\frac{1}{a}+\frac12\Bigl[r_e+\frac{1}{a\,m^{2}}\Bigr]k^{2}+\dots,\qquad k\ll m .
\end{equation}
The second term in the bracket comes from that factor alone and is
small near the resonance, $|1/a|\ll m^{2}|r_e|$.  The effective range of the
phase is therefore $r_e+1/(am^{2})$; at the resonance, $1/a=0$, the two
coincide.  The number $r_e$ itself does not depend
on the coupling: it is a property of the channel, and it is what we call the
effective range.  At a given mass the scale alone fixes it.  For $\lambda$
above $\tfrac12e\,m$ it is negative, and its magnitude grows as the logarithm
of the scale.  If the integral is stopped at an upper limit $\Lambda$ instead, the
range is $(4/\pi m)[E_\Lambda/\Lambda-\operatorname{arsinh}(\Lambda/m)]$, which
is that of the scale $\lambda=\Lambda$ plus $m/(\pi\Lambda^{2})$ and terms of
higher order in $1/\Lambda^{2}$.

The two roots of the quadratic in \eqref{eq:Dexp} approximate the two poles of
Fig.~\ref{fig:pole}, the better the larger the logarithm of the scale.  They
meet at $q=1/r_e$, which estimates the point $q_*$
where the virtual state leaves the real axis.  For a nonrelativistic contact
interaction the effective range is not negative: it is $4/(\pi\Lambda)$ at an
upper limit $\Lambda$ and vanishes when the limit is removed.  The loop then
has no maximum on the side $q<0$, and a virtual state exists at every negative
scattering length: the point $q_*$ is the signature of the negative effective
range of the channel.

\subsection{The effective range from the circle}

The figure of Sec.~\ref{sec:levels} gives the expansion \eqref{eq:Dexp} directly.  At
$\Delta=0$ and $\mu<m$ the apex lies on the base, between the points
$\pm E_p$.  The ellipse through it is the segment between its foci, so that
$h_E=0$ and $\tfrac12S=E_p$, and the height of the circle above $\mu$ is
\begin{equation}\label{eq:twobody}
  h_\mu^{2}=p^{2}+q^{2},\qquad q^{2}=m^{2}-\mu^{2},\qquad
  \frac{\partial S}{\partial(\Delta^{2})}\bigg|_{\Delta=0}=\frac{S/2}{h_\mu^{2}}=\frac{E_p}{p^{2}+q^{2}} .
\end{equation}
The square of the height above $\mu$ is the denominator of the loop, and at
$p=0$ the height is $q$ itself: the binding momentum of a pair is the height
of the smallest circle above the foot of the apex.  For the point of the
circle above $\mu$, $(\tfrac12S)^{2}=\mu^{2}+h_\mu^{2}$, and the loop up to
$\Lambda$ splits in two,
\begin{equation}\label{eq:spinors}
  V(\Lambda,q)=\int_0^{\Lambda}\frac{p^{2}\,dp}{E_p}+\mu^{2}\int_0^{\Lambda}\frac{p^{2}\,dp}{E_p\,h_\mu^{2}}
  =\int_0^{\Lambda}\frac{p^{2}\,dp}{E_p}+\Bigl(\frac{E_{\rm pair}}{2m}\Bigr)^{2}V_s(\Lambda,q),\qquad
  V_s(\Lambda,q)=\int_0^{\Lambda}\frac{m^{2}}{E_p}\,\frac{p^{2}\,dp}{p^{2}+q^{2}},
\end{equation}
with $E_{\rm pair}=2\mu=2\sqrt{m^{2}-q^{2}}$, the energy of the pair.  The
first sum does not depend on that energy and cancels in the difference
$V(q_B)-V(q)$.  The second carries the energy through its factor $\mu^{2}$.  At the threshold the
height above $\mu$ is the momentum itself, the measure cancels against it, and
what is left is the sum of the inverse radii, $\int dp/E_p$.  It grows as a
logarithm because the radius grows only linearly with the momentum.  As the
pair binds deeper, $\mu^{2}$ falls by $q^{2}$, and this puts the logarithm into
the coefficient of $q^{2}$ in $V$.  This coefficient is the normalization of
the field of the pair of Sec.~\ref{sec:crossover}, the response of the loop to
$\mu^{2}$,
\begin{equation}\label{eq:Zfield}
  Z(q)\equiv\frac{\partial V}{\partial(\mu^{2})}=\int_0^{\Lambda}\frac{(S/2)\,p^{2}\,dp}{h_\mu^{4}}
  =\frac{\pi m}{4q}+\operatorname{arsinh}\frac{\Lambda}{m}-\frac{E_\Lambda}{\Lambda}+O(q),\qquad
  r_e=-\frac{4}{\pi m}\lim_{q\to0}\Bigl[Z(q)-\frac{\pi m}{4q}\Bigr] ,
\end{equation}
where the first term is the threshold of two free particles and $r_e$ is the
range at the upper limit $\Lambda$.  With all momenta integrated the two terms
after the first are $\ln(2\lambda/m)-1$, and $r_e$ is that of
\eqref{eq:Dexp}.  A negative effective range is a positive normalization of
the field of the pair.

\subsection{The closed channel}\label{sec:closed}

A potential well of radius $R$ has an effective range of the order of $R$, and
a contact interaction of nonrelativistic particles has $r_e=4/(\pi\Lambda)$ at
an upper limit $\Lambda$, which vanishes when all momenta are integrated.  A
negative effective range much longer than the range of the force is the mark
of a pair that spends part of its time in another state, a closed channel, as
at a Feshbach
resonance~\cite{Petrov2004,GurarieRadzihovsky2007,ChinEtAl2010}.

No closed channel has been put in here.  Its part is played by the pair itself
at relative momenta far above the mass, where a mode holds the charge
$\mu\Delta^{2}/p^{3}$ of Sec.~\ref{sec:arc}.  These modes carry the logarithmic
term of the density \eqref{eq:Qinf}, which is measured from the scale
$\lambda$, and seen from $k\ll m$ they act as a second channel.  In the model itself the pair
is composite in the sense of Sec.~\ref{sec:crossover}, and the closed channel
belongs to its low-energy description.

The mean-field crossover with a finite effective range is that of the
two-channel model of a Feshbach resonance~\cite{GurarieRadzihovsky2007}.  In a
boson--fermion model the effective range comes from the coupling of the boson
to the fermions~\cite{DengSchmittWang2007}, and Ref.~\cite{Braun2012}
discusses the narrow resonance as the limit of a small boson--fermion
coupling.  For two spin states the density is
$n=Q_\mu/\pi^{2}=k_F^{3}/3\pi^{2}$, and with this normalization the gap
equation and the density of the two-channel model are
\begin{equation}\label{eq:twochannel}
  \begin{gathered}
  \int_0^{\infty}\Bigl[\frac{k^{2}}{2E_k}-m\Bigr]dk+\frac{\pi m}{2}\Bigl[\frac{1}{a}-r_e\,m\,\mu_{\rm nr}\Bigr]=0,\\
  \tfrac13k_F^{3}=\int_0^{\infty}k^{2}v_k^{2}\,dk+\frac{\pi}{4}\,m^{2}|r_e|\,\Delta^{2},\qquad
  E_k=\sqrt{\bigl(k^{2}/2m-\mu_{\rm nr}\bigr)^{2}+\Delta^{2}} .
  \end{gathered}
\end{equation}
Here $\mu_{\rm nr}\equiv\mu-m$, and
$v_k^{2}=\tfrac12[1-(k^{2}/2m-\mu_{\rm nr})/E_k]$ is the occupation of a mode.
The last term is the density of the fermions bound in closed-channel
molecules, twice the density of the molecules; with $r_e=0$ these are the
equations of Ref.~\cite{MariniPistolesiStrinati1998}.

Let $k_F/m\to0$ with $k_Fr_e$ fixed, so that $\lambda$ is exponentially large.
Then the density of \eqref{eq:Qinf} and the gap equation in the form
$D|_{q_B=0}+\pi m/2a=0$ of Sec.~\ref{sec:crossover} become
\eqref{eq:twochannel}.  Together with a constant from $q_\mu$, the logarithmic
term of the density,
$\mu\Delta^{2}\ln(2\lambda/\ell_0)\to m\Delta^{2}\ln(2\lambda/m)$, is the part
of the density that the two-channel model assigns to the closed channel,
$\tfrac14\pi\,m^{2}|r_e|\Delta^{2}$.  Together with a constant from $D_0$, the
logarithmic term of the gap equation is the dependence of the coupling on the
energy, $-\tfrac12\pi\,m\,r_e\,m\mu_{\rm nr}$.  The relativistic crossover at
$k_F\ll m$ is thus that of a resonance of finite width.  The resonance is
narrow in the sense that matters for the crossover,
$\gamma_s=(8/\pi)/(k_F|r_e|)\ll1$ in the notation of
Ref.~\cite{GurarieRadzihovsky2007}, only when $\ln(2\lambda/m)\gg2m/k_F$.  At a
fixed $\lambda$ the product $k_Fr_e$ vanishes as $k_F/m\to0$, and the crossover
tends to that of Ref.~\cite{MariniPistolesiStrinati1998}.

\begin{figure}[t]
  \centering
  \includegraphics[width=0.96\linewidth]{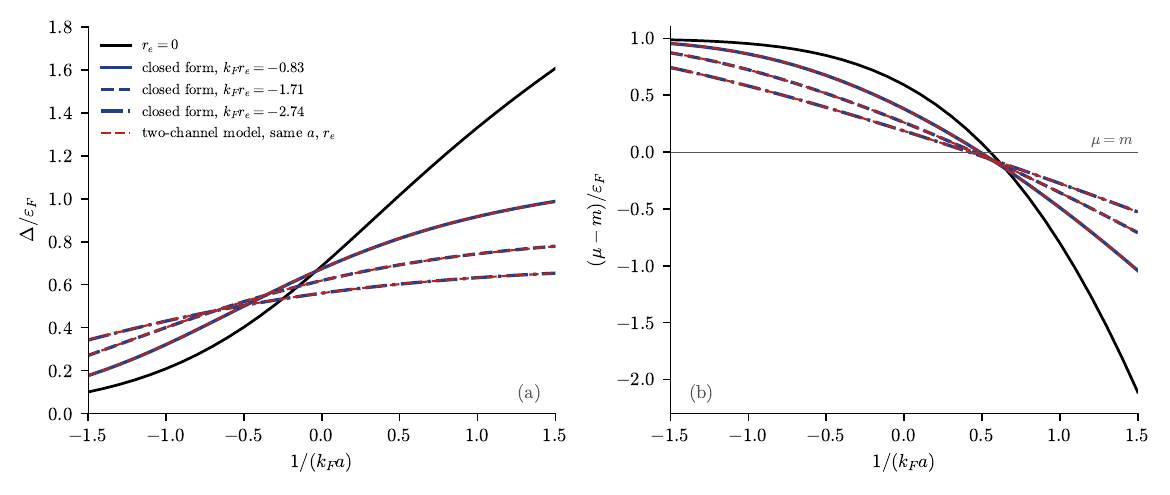}
  \caption{The crossover at $k_F=0.05\,m$ and three values of $k_Fr_e$,
  $-0.83$, $-1.71$ and $-2.74$:
  (a)~the gap and (b)~the chemical potential measured from the mass, in units
  of $\varepsilon_F=k_F^{2}/2m$, against the inverse scattering length
  $1/(k_Fa)$.  Blue: the equations
  \eqref{eq:Qinf}, with all momenta integrated.  Red dashes: the two-channel
  model \eqref{eq:twochannel} with the same $a$ and $r_e$.  Black: $r_e=0$,
  the crossover of Ref.~\cite{MariniPistolesiStrinati1998}.}
  \label{fig:crossover}
\end{figure}

Figure~\ref{fig:crossover} shows the limit of a fixed $k_Fr_e$ at
$k_F=0.05\,m$, for three values of $k_Fr_e$; the energies are in units of
$\varepsilon_F=k_F^{2}/2m$.  The gap and the chemical potential of
\eqref{eq:Qinf} coincide with those of the two-channel model with the same $a$
and $r_e$ to within $2\times10^{-3}\,\varepsilon_F$ over the whole range of the
figure~\cite{supp}, and both are far from the crossover of $r_e=0$: in this
limit the two give the same mean-field equations, term by term, and the same
solutions.  The effective range weakens the dependence on the scattering
length: at weak attraction the gap is larger than at $r_e=0$ and the chemical
potential lower, and at strong attraction the gap is smaller and the chemical
potential higher.  At this $k_F$ these values of $k_Fr_e$ require
$\lambda/m=6\times10^{5}$, $6\times10^{11}$ and $6\times10^{18}$: the figure
shows the limit, in which $\lambda$ is exponentially large.

\subsection{Which pairs have the logarithm}

The negative effective range comes from the pairing vertex; relativity alone
does not give it.  The integrand of the loop \eqref{eq:Vdef} is $E_p^{2}/E_p$
times $p^{2}/(p^{2}+q^{2})$, and its numerator $E_p^{2}$ is, up to the factor
$8$, the squared matrix element of the pairing vertex between a
fermion at $\mathbf p$ and an antifermion at $-\mathbf p$, summed over the
spins.  For three
vertices, and for two fermions paired by the charge conjugation
$C=i\gamma^{2}\gamma^{0}$, it is
\begin{equation}\label{eq:vertices}
  \sum_{\rm spins}\bigl|\bar u(\mathbf p)\,\Gamma\,v(-\mathbf p)\bigr|^{2}=
  \begin{cases}
    8p^{2}, & \Gamma=1,\\
    8E_p^{2}, & \Gamma=i\gamma_5,\\
    8m^{2}, & \Gamma=\gamma^{0}\gamma_5,
  \end{cases}
  \qquad
  \sum_{\rm spins}\bigl|\bar u(\mathbf p)\,\Gamma C\,\bar u(-\mathbf p)^{\rm T}\bigr|^{2}=
  \begin{cases}
    8p^{2}, & \Gamma=1,\\
    8E_p^{2}, & \Gamma=\gamma_5 .
  \end{cases}
\end{equation}
The two columns agree because the spinors $C\bar u^{\rm T}$ are the spinors
$v$ of the antifermion.  The pairing term of \eqref{eq:H}, a unit Dirac matrix
in the basis used there, is the vertex $\gamma_5$ between the fermion and the
antifermion, and its matrix element, $8E_p^{2}$, grows with the energy.  The
scalar vertex vanishes at the threshold as $p^{2}$: it pairs in a $p$ wave.
With the vertex $\gamma^{0}\gamma_5$ the loop would be $V_s$ of
\eqref{eq:spinors} itself.  Its logarithm does not depend on the energy, so
its effective range has no logarithm: it tends to $4/\pi m$ for any rule that stops
the momentum, as for a relativistic contact interaction of spinless particles
at resonance~\cite{Pascalutsa2016}.  Two fermions paired by
$C\gamma_5$, as in the scalar diquark of quark matter, have the loop of the
pair of \eqref{eq:H}: at the same $m$ and $\lambda$ the two have the same
$r_e$, and only their scattering lengths differ.

\subsection{The nonrelativistic corner and earlier solutions}

In one corner of the parameter space the answer has long been known.  Let
$m\to\infty$ with $\mu_{\rm nr}$ and $\Delta$ fixed.  The level of the
antiparticle then moves away to $2m$, and the problem reduces to the problem
with one point of Sec.~\ref{sec:levels}.  The modulus and the parameter of the third
kind tend to one modulus, that of Ref.~\cite{MariniPistolesiStrinati1998}, the
arc grows to a quarter of the ellipse, and the integral of the third kind is
no longer a new function.  Only the complete integrals of the first and second
kinds remain, and the two equations become the gap equation and the
density of that reference (Appendix~\ref{app:corner}).  This is the corner at
a fixed ratio $\lambda/m$: the logarithm of \eqref{eq:Qinf} enters it with
corrections of the order $k_Fr_e$, which vanish there.  At a fixed $k_Fr_e$
the corner is the two-channel model \eqref{eq:twochannel}, as in
Sec.~\ref{sec:closed}.

Away from the corner the two equations \eqref{eq:Qinf} are solved as they
stand; the curves at $k_F=m/2$ are in the Supplemental
Material~\cite{supp}.  There, too, the larger $\lambda$, the larger $|r_e|$
and the weaker the dependence of the gap on the scattering length.

Earlier finite-cutoff calculations provide useful benchmarks for the present
formulation.  In Ref.~\cite{HeZhuang2007} the crossover was solved at
$\Lambda=10\,m$; up to corrections of the order $m^{2}/\Lambda^{2}$, this
cutoff corresponds to the scale $\lambda=10\,m$.
Reference~\cite{Brauner2008} considered $k_F=0.2\,m$ and $\Lambda=5\,m$,
corresponding to $k_Fr_e=-0.33$.  When $\lambda$ is chosen to reproduce the
same effective range, the equations with all momenta integrated recover the
zero-temperature gap of the equations of Ref.~\cite{Brauner2008}, with its
cutoff, to within $0.2$ per cent throughout the coupling range shown in its
Fig.~2~\cite{supp}.  This agreement provides a nontrivial cross-check of the
closed form and of the mapping between $\lambda$ and the effective range.

\section{Quark matter in the Nambu--Jona-Lasinio model}\label{sec:quark}

The Nambu--Jona-Lasinio model of quark matter has the two levels
\eqref{eq:levels} in more than one of its phases.  At a nonzero isospin
chemical potential the pair is the pion, and in the two-flavour colour
superconductor it is the scalar diquark.  In this model the scale $\lambda$ is
not a free parameter: the pion decay constant fixes it, for the pion and for
the diquark alike.  With a second condensate the figure has a second apex, and
the closed form no longer applies.

\subsection{Pion condensation}\label{sec:pion}

In the model of Ref.~\cite{companion} the fermion is a quark,
$\mu=\mu_I/2$ is half the isospin chemical potential, the gap is the charged
pion condensate, and the mass $M$ of the quark is itself a condensate.  With
$\rho^{2}=M^{2}+\Delta^{2}$ and all momenta integrated, its potential is
\begin{equation*}
  \Omega=\tfrac12A'\rho^{2}-B'M-\frac{N_c}{\pi^{2}}\,I_\lambda(M,\mu,\Delta) ,
\end{equation*}
where $B'$ is proportional to the current mass of the quarks.  The coupling,
which stands in $A'$, has taken the power $\rho^{2}\Lambda^{2}$ whole, so that
no part of the logarithm is absorbed and the whole coefficient $\tfrac12c_3$
comes with the one scale $\lambda$.  The mass and the gap follow from
$\partial\Omega/\partial M=0$ and $\partial\Omega/\partial\Delta=0$, and the
three numbers $A'$, $B'$ and $\lambda$ are fixed in the vacuum.

\subsubsection{The pion}

In this model the scale $\lambda$ is fixed by the pion decay constant,
$f_\pi^{2}=(N_cM^{2}/2\pi^{2})[\ln(2\lambda/M)-1]$~\cite{companion}, and the
quark--antiquark channel of the pion has
\begin{equation}\label{eq:reNJL}
  r_e=-\frac{8\pi f_\pi^{2}}{N_cM^{3}}=-\frac{8\pi}{N_cM\,g_{\pi qq}^{2}}\approx-0.50\ \text{fm},
  \qquad g_{\pi qq}=\frac{M}{f_\pi},
\end{equation}
for $M=303.5$~MeV and $f_\pi=92.4$~MeV, the values of Ref.~\cite{companion};
$g_{\pi qq}=M/f_\pi$ is the coupling of the pion to the quarks in the chiral
limit.  This is the effective range that the channel would have at resonance.
The second form is the effective range
$-8\pi/(m^{2}g_s^{2})$ of the two-channel model of
Ref.~\cite{GurarieRadzihovsky2007}, with the pion as the molecule and
$g_s^{2}=N_cg_{\pi qq}^{2}/M$ as its coupling to a quark and an antiquark at
the threshold.  At this mass $|r_e|=0.78/M$ is shorter than
the Compton wavelength of the quark, so the channel is not close to a narrow
resonance, and $r_e$ does not measure a closed-channel admixture of the pion.
The pion itself is deeply bound, $q_B\approx M$, and the effective range of its
phase, $r_e+1/(aM^{2})$, has the other sign.

The two forms of \eqref{eq:reNJL} are values of one function, the
normalization \eqref{eq:Zfield} of the field of the pair, at different
energies of the pair.  As a function of the momentum $q$,
\begin{equation*}
  Z(q)=\int_0^{\Lambda}\frac{E_p\,p^{2}\,dp}{(p^{2}+q^{2})^{2}}=-\frac{\partial V}{\partial(q^{2})} .
\end{equation*}
At the threshold of two quarks, $q\to0$, it gives the effective range,
$r_e=-(4/\pi M)K$ with $K=\lim_{q\to0}[Z(q)-\pi M/4q]$, which with all momenta
integrated is $\ln(2\lambda/M)-1$.  At zero energy, $q=M$, it is
$Z(M)=\int_0^{\Lambda}p^{2}dp/E_p^{3}$, the integral of $f_\pi$ in
Ref.~\cite{companion}, $f_\pi^{2}=(N_cM^{2}/2\pi^{2})\,Z(M)$.  At a vanishing
gap this is also the coefficient of $\mu^{2}\Delta^{2}$ in
\eqref{eq:integral}: at any gap
$S=2E_\rho+\mu^{2}\Delta^{2}/E_\rho^{3}+O(\mu^{4})$ with
$E_\rho^{2}=p^{2}+M^{2}+\Delta^{2}$, so that the coefficient is the same
integral with $\rho$ in place of $M$.  The scale that $f_\pi$ fixes is thus
the scale of the logarithm that the chemical potential brings.  With all
momenta integrated $Z(M)=K$.  The same number thus appears at two places of
the curve of the loop: in units of the mass, $M\,r_e=-(4/\pi)K$ is its
curvature at the threshold, where Fig.~\ref{fig:pole} shows it, and its slope
at the end $q=M$, beyond the frame of that figure, where the pion of the
chiral limit sits.  At the
pole of the pion, $q=q_B$ with $2(M^{2}-q_B^{2})^{1/2}=m_\pi$, $Z$ gives the
residue of the amplitude of a quark and an antiquark, and the coupling of the
pion to the quarks read off the residue is
$g_{\pi qq}^{2}=2\pi^{2}/[N_cZ(q_B)]$.  By \eqref{eq:Vdef} this pole is the
onset of the pion phase, $2\mu=\mu_I=m_\pi$.  In the chiral limit it is at
zero energy, $q_B=M$: the coupling is $g_{\pi qq}=M/f_\pi$, the relation of
Goldberger and Treiman, and the two forms of \eqref{eq:reNJL} coincide.
Off the chiral limit the first form holds at any current mass of the quarks,
and the second, with the coupling read at the pole, depends on it~\cite{supp}.

\subsubsection{The constituent quark mass}

The vacuum does not fix the constituent mass of the quarks.  The three numbers
$A'$, $B'$ and $\lambda$ of the potential are fixed by three quantities of the
vacuum, $M$, $f_\pi$ and $m_\pi$~\cite{companion}, and of these only $M$ is
not an observable.  At a given $f_\pi$ the formula for $f_\pi$ and the first form of
\eqref{eq:reNJL} fix the scale and the effective range as functions of the
mass,
\begin{equation*}
  \lambda=\frac{M}{2}\exp\Bigl[1+\frac{2\pi^{2}f_\pi^{2}}{N_cM^{2}}\Bigr],\qquad
  r_eM^{3}=-\frac{8\pi f_\pi^{2}}{N_c} ,
\end{equation*}
so that the magnitude of the effective range falls as the inverse cube of the
mass.

\begin{figure}[t]
  \centering
  \includegraphics[width=\linewidth]{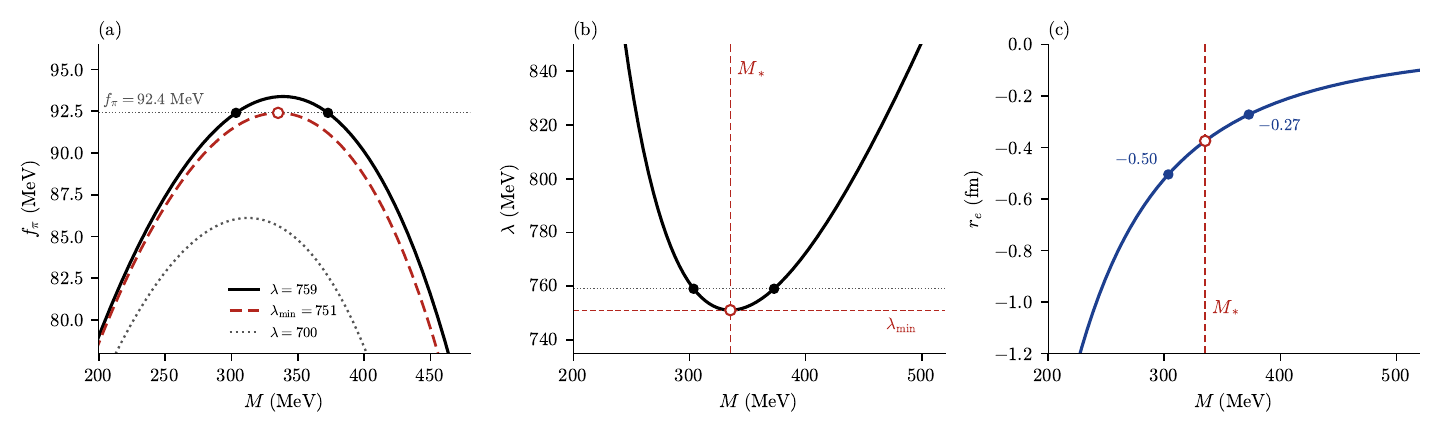}
  \caption{The constituent mass of the quarks as the free number, for $N_c=3$; masses and
  scales in MeV.  (a)~The pion decay constant against the mass at three fixed
  scales: $\lambda=759$ (solid), $\lambda_{\min}=751$ (dashed) and $700$
  (dotted).  The horizontal line is $f_\pi=92.4$~MeV: the first curve meets it
  at two masses, $303.5$ and $373$~MeV (dots), the second touches it at
  $M_*=335$~MeV (open circle), and the third stays below it.  (b)~The scale
  against the mass at $f_\pi=92.4$~MeV, with its minimum $\lambda_{\min}$ at
  $M_*$; the dots are the two masses of (a).  (c)~The effective range
  \eqref{eq:reNJL}, in fm, at the same $f_\pi$.}
  \label{fig:knob}
\end{figure}

At a fixed $\lambda$ the pion decay constant, as a function of the mass,
rises, passes a maximum and falls (Fig.~\ref{fig:knob}a),
\begin{equation*}
  \frac{df_\pi^{2}}{dM}=\frac{N_cM}{2\pi^{2}}\,(2K-1),\qquad K=\ln\frac{2\lambda}{M}-1 .
\end{equation*}
The maximum is at $K=\tfrac12$, that is at $M=2\lambda e^{-3/2}$, and its
height is $f_\pi=\sqrt{N_c}\,M/2\pi$.  The measured $f_\pi$ cannot exceed this
height, and this bounds the scale from below,
\begin{equation*}
  \lambda\ge\lambda_{\min}=\frac{e^{3/2}\,\pi f_\pi}{\sqrt{N_c}},\qquad
  M_*=\frac{2\pi f_\pi}{\sqrt{N_c}} .
\end{equation*}
At $\lambda=\lambda_{\min}$ the measured $f_\pi$ is reached at one mass,
$M_*$.  At a larger $\lambda$ it is reached at two masses, one below $M_*$ and
one above (Fig.~\ref{fig:knob}b), and the two have different effective ranges
(Fig.~\ref{fig:knob}c).  A value of $\lambda$ therefore does not single out
the model, and a value of $M$ does: the free number is the mass.  At the
maximum the coupling of Goldberger and Treiman is
$g_{\pi qq}=M/f_\pi=2\pi/\sqrt{N_c}$, and $M_*$ is the mass of the quarks
found in the linear sigma model generated at one
loop~\cite{DelbourgoScadron1995}.

The scale is thus fixed once, in the vacuum: the measured $f_\pi$ gives
$\lambda$ at a given $M$, and with it the effective range of the channel of
the pion.  The mass $M$, the constituent mass of the quark in the vacuum, is
the one number that the vacuum leaves free.

\subsubsection{Isospin density}

\begin{figure}[t]
  \centering
  \includegraphics[width=0.96\linewidth]{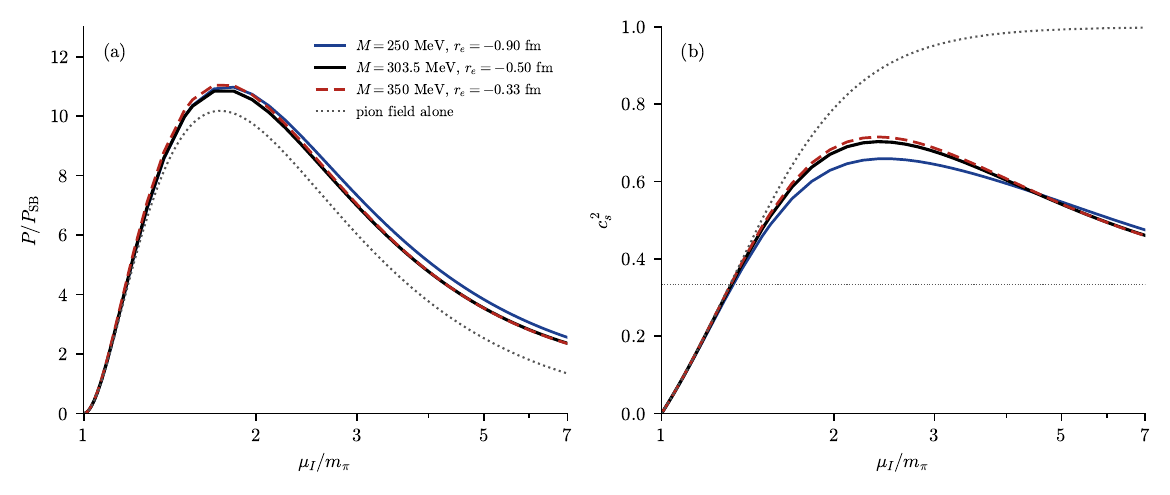}
  \caption{Pion condensation in the model of Ref.~\cite{companion} with all
  momenta integrated: (a)~the pressure in units of
  $P_{\rm SB}=\mu_I^{4}/32\pi^{2}$ and (b)~the squared speed of sound, against
  $\mu_I/m_\pi$, at the physical $f_\pi=92.07$~MeV and $m_\pi=139.57$~MeV and
  three masses of the quarks.  The effective range in the legend is that of
  \eqref{eq:reNJL}.  Dotted: the pion field alone, the leading order of chiral
  perturbation theory.  The thin line in (b) is the conformal value
  $\tfrac13$.}
  \label{fig:isospin}
\end{figure}

At an isospin chemical potential above the mass of the pion the model is the
crossover of this paper with the charged pion as the pair, and with a mass of
the quark that follows, with the gap, from the two equations of the potential.
Figure~\ref{fig:isospin} shows the pressure and the squared speed of sound,
with all momenta integrated, from the onset of the condensation,
$\mu_I=m_\pi$, to $7\,m_\pi$, at one $f_\pi$ and three masses of the quarks.
At a fixed $f_\pi$ the equation of state depends only weakly on the mass that
the vacuum leaves free.  Between $1.5$ and $7\,m_\pi$ the three pressures
differ by less than 10 per cent,
although the effective range \eqref{eq:reNJL} changes almost threefold, while
a change of $f_\pi$ moves the pressure much more~\cite{supp}.  The squared
speed of sound has a peak above the conformal value $\tfrac13$, and the height
of the peak grows with the mass, from $0.66$ to $0.72$.  Here, however, the
effective range cannot be read as in Sec.~\ref{sec:range}.  Near the onset the
density is low, but the pion is deeply bound, far from the resonance; farther
on the mass of the quark falls, and the density is no longer low on its
scale.  The crossover is the full one, which the closed form describes and
the two-channel model does not.  Lattice QCD has computed the equation of
state of this matter at zero temperature~\cite{NPLQCD2025}; we make no
comparison with it here.

\subsection{Colour superconductivity}\label{sec:csc}

In the two-flavour colour superconductor the paired quarks have the two
levels \eqref{eq:levels}, with the chemical potential of the quarks in place
of $\mu$~\cite{Shovkovy2005}, and the closed form applies to them as it
stands.  The pair is the scalar diquark, two quarks paired by $C\gamma_5$.  By
\eqref{eq:vertices} its loop in the vacuum is the loop of the pair of
\eqref{eq:H}, so that the diquark channel has the logarithm of the pion
channel and, at the same $M$ and $\lambda$, its effective range.  With the
mass as a parameter this is the model of Ref.~\cite{Brauner2008}, in which the
quarks of two colours pair and those of the third are a free gas; its
equations at zero temperature were compared with the closed form in
Sec.~\ref{sec:range}.

Two renormalization-group studies of this phase allow us to check the closed
form against results obtained independently.
Reference~\cite{GholamiHofmannBuballa2025} constructed a treatment of the
phase in the Nambu--Jona-Lasinio model that is consistent under the
renormalization group: the vacuum part is regularized and fitted, and the
medium part is integrated to infinity.  A logarithmic divergence then remains
in the medium, $-(2/\pi^{2})\mu^{2}\Delta^{2}\ln\Lambda$, which the authors
cancel by counterterms that vanish in the vacuum, in three schemes, massive,
massless and minimal, each consistent under the renormalization group.  This
is the term $\mu^{2}\Delta^{2}$ of the coefficient \eqref{eq:Sasym}, taken
with the number of the paired quarks, so that their result confirms the
coefficient of the logarithm of Sec.~\ref{sec:crossover} in the medium.

In each of their schemes the factor that multiplies
$-(2/\pi^{2})\mu^{2}\Delta^{2}$ in the potential is the integral of
$p^{2}/E_\rho^{3}$, the coefficient of $\mu^{2}\Delta^{2}$ of
Sec.~\ref{sec:pion}, which we write $K_\rho$.  With their cutoff $\Lambda'$
and $\rho^{2}=M^{2}+\Delta^{2}$,
\begin{equation*}
  \begin{gathered}
  K_\rho^{\rm massive}=\operatorname{arsinh}\frac{\Lambda'}{\rho}-\frac{\Lambda'}{\sqrt{\Lambda'^{2}+\rho^{2}}},\qquad
  K_\rho^{\rm massless}=\operatorname{arsinh}\frac{\Lambda'}{\Delta}-\frac{\Lambda'}{\sqrt{\Lambda'^{2}+\Delta^{2}}}+\ln\frac{\Delta}{\rho},\\
  K_\rho^{\rm minimal}=\ln\frac{2\Lambda'}{\rho}-1 .
  \end{gathered}
\end{equation*}
With all momenta integrated the factor is
\begin{equation*}
  K_\rho=\ln\frac{2\lambda}{\rho}-1 ,
\end{equation*}
and at $\lambda=\Lambda'$ their three factors agree with it up to
$3\rho^{2}/4\Lambda'^{2}$, $3\Delta^{2}/4\Lambda'^{2}$ and zero~\cite{supp}.
The schemes, built independently of the closed form, share its logarithm and
its scale, and the terms in which they differ at a finite $\Lambda'$ are of
the order $\rho^{2}/\Lambda'^{2}$.  In the vacuum, where $\rho=M$, $K_\rho$ is
the $K$ of Sec.~\ref{sec:pion}.

Reference~\cite{GholamiEtAl2026} renormalizes a quark--meson--diquark model and
gives a second check.  Its constant $\tilde Z_\Delta$ is $K$ times
$g_\Delta^{2}/2\pi^{2}$, where $g_\Delta$ is the coupling of the diquark to
the quarks.  Since $K$ gives the effective range of the channel of two quarks,
$r_e=-(4/\pi M)K$ by Sec.~\ref{sec:range}, with all momenta integrated
\begin{equation*}
  r_e^{qq}=-\frac{8\pi\,\tilde Z_\Delta}{M\,g_\Delta^{2}} ,
\end{equation*}
the effective range of the pion channel at the same $\lambda$, as
\eqref{eq:vertices} requires.  That work has a fourth scheme, vacuum
matching, which takes the counterterm at the vacuum condensate.  Carried over
to the Nambu--Jona-Lasinio model, it is the potential with all momenta
integrated, with $\lambda$ fixed by $f_\pi$: the renormalization-group
construction arrives at the closed form.

For the pion $K$ is set by $f_\pi^{2}$, a quantity of the vacuum
(Sec.~\ref{sec:pion}).  If the vacuum part and the medium part are integrated
together, over all momenta, there is one scale for both channels, and $f_\pi$
sets $K$ for the diquark as well.  At three colours no quantity of the vacuum
fixes the factor of the diquark channel separately; at two colours the
symmetry of Pauli and G\"ursey ties the two
channels~\cite{KhunjuaKlimenkoZhokhov2020}.  The colour superconductor considered here has two flavours and
is not neutral.  In neutral matter of three flavours the pairs $us$ and $ds$
are made of quarks of different masses; their levels are not those of
\eqref{eq:levels}, and the closed form does not apply.

\subsection{More than one apex}\label{sec:pairs}

The figure also shows where the closed form ends.  A second condensate with a
chemical potential of its own is a second apex over
the same base.  In \eqref{eq:H} the terms of the momentum, of the mass and of
the gap anticommute, and the charge of the chemical potential commutes with
the first two and anticommutes with the third.  We call a condensate and a
chemical potential conjugate when the charge of that chemical potential
anticommutes with the term of the condensate and commutes with every other
term.  A conjugate couple is an apex, $w_j=\mu_j+i\Delta_j$; the mass has no
chemical potential of its own and stays in the base.  For $n$ apexes over one base the circle of
diameter $S$ stands at the height $h_E$ above the end $E_p$ of the base and at
a height $h_j$ above the foot of each apex, and \eqref{eq:ell2} becomes
\begin{equation}\label{eq:pairlaw}
  \begin{gathered}
  \frac{S^{2}}4-h_E^{2}=p^{2}+m^{2},\qquad
  \sum_{j=1}^{n}\frac{\Delta_j^{2}}{h_j^{2}}=\frac{4h_E^{2}}{S^{2}},\qquad
  h_j^{2}=\frac{S^{2}}4-\mu_j^{2},\\
  \eta=\pm r_1\pm r_2\pm\dots\pm r_{n+1} .
  \end{gathered}
\end{equation}
For one apex the second equation is the law of the two heights
\eqref{eq:heights}.  In general it is an equation of degree $n+1$ for $S^{2}$
with real positive roots, and with $r_k$ half the square root of the $k$-th
root the levels $\eta$ of the mode are all the sums in the second line; the
largest root is the circle, $r_1=\tfrac12S$.  The reason is that such a
Hamiltonian is a bilinear form in $2n+2$ anticommuting matrices of unit
square: its levels are the sums of the singular values of the matrix of its
coefficients, taken with all signs, and the $r_k$ are these singular
values~\cite{supp}.

As long as $r_1$ exceeds the sum of the other $r_k$ at every momentum, no
level crosses zero, and the potential integrates $S$ as before.  The
integration by parts of Sec.~\ref{sec:arc} then carries over word for word,
with the boundary term of \eqref{eq:oneline} as it stands and
$\rho^{2}=m^{2}+\sum_j\Delta_j^{2}$.  The integral grows by the same two
powers and by one logarithm,
\begin{equation}\label{eq:survn}
  S=2p+\frac{\rho^{2}}{p}+\frac{c_3/2}{p^{3}}+O(p^{-5}),\qquad
  c_3=2\sum_{j=1}^{n}\mu_j^{2}\Delta_j^{2}-\frac{\rho^{4}}{2} ,
\end{equation}
to whose coefficient each apex adds its $\mu_j^{2}\Delta_j^{2}$.  If a level
crosses zero in a window of momenta, the integrand differs from $S$ there; the
growing terms come from large momenta and stay the same.  What does
not carry over is the last step.  By \eqref{eq:pairlaw} the square of the
momentum is a rational function of $S^{2}$, and $p^{2}S\,dp$ is a differential
on a curve of genus $n$.  With two apexes the integrals that are left are
hyperelliptic, and for parameters in general position they do not reduce to
elliptic ones.  The closed form of this paper is the case of one apex.

\section{Conclusions}\label{sec:conclusions}

The sum of the two levels of a Dirac fermion with a gap, integrated over the
momentum, is the one integral of its mean-field potential at zero temperature.
We have read its closed form off a geometric figure.  The two levels are the
distances from the point $w=\mu+i\Delta$ to the points $\mp E_p$, their sum is
the diameter of a circle whose two heights obey
$h_E\,h_\mu=\tfrac12S\Delta$, and the momentum line is a finite arc of the
ellipse of $p=0$.  Along this arc the charge, the scalar density and the pair
amplitude, summed over the modes, are combinations of Legendre's three
integrals, and only the integral of the third kind grows.  As the upper limit
grows, the integral grows by two powers of the limit, which are absorbed into
a constant and into the coupling, and by one logarithm.  With all momenta
integrated what is left of it is
\begin{equation*}
  I_\lambda(m,\mu,\Delta)=\mathcal C(m,\mu,\Delta)+\frac{c_3}{2}\ln\frac{2\lambda}{\ell_0},
  \qquad \frac{c_3}{2}=\mu^{2}\Delta^{2}-\frac{\rho^{4}}{4} ,
\end{equation*}
with the finite part \eqref{eq:Cfin} and a scale $\lambda$ that the integral
does not fix.

The gap equation and the density of the relativistic crossover are
derivatives of this formula, \eqref{eq:Qinf}, and solving the crossover means
finding the common root of two closed expressions.  The equations themselves
are those of
Refs.~\cite{HeZhuang2007,Brauner2008}, where a cutoff is kept as a parameter.
In the closed form that parameter is the scale $\lambda$, and it has a
meaning: it fixes the effective range of the channel of the pair,
$r_e=-(4/\pi m)[\ln(2\lambda/m)-1]$.  The effective range is negative, and
its magnitude grows as the logarithm of the scale, because the matrix element
of the pairing vertex $\gamma_5$ grows with the energy; relativity alone does
not give it.
At $k_F\ll m$ and a fixed $k_Fr_e$ the crossover is that of the two-channel
model of a Feshbach resonance with the same scattering length and effective
range, and the part of the closed channel is played by the pair itself at
momenta far above the mass.

In the Nambu--Jona-Lasinio model the scale is fixed once, in the vacuum, by
the pion decay constant.  The effective range of the pion channel is then
$-8\pi f_\pi^{2}/(N_cM^{3})$, and the mass $M$ of the quark in the vacuum is
the one number that stays free.  The renormalization-group studies of the
two-flavour colour superconductor~\cite{GholamiHofmannBuballa2025,GholamiEtAl2026}
find the same logarithm in the medium, and their schemes reproduce the factor
$\ln(2\lambda/\rho)-1$ of the closed form at $\lambda=\Lambda'$ up to terms of
the order $\rho^{2}/\Lambda'^{2}$.

It would be interesting to look for a negative effective range of this kind
elsewhere: in lattice computations of four-fermion models, where its magnitude
should grow as the logarithm of the inverse lattice spacing; in atomic gases,
where a large negative effective range has been realized at a narrow Feshbach
resonance~\cite{Hazlett2012}; and in Dirac materials, which have the same two
levels~\cite{MizoguchiOgata2015,Hayasaka2026}.  Our estimates of the effect on
a lattice and in atomic gases are in the Supplemental Material~\cite{supp}.

The closed form is a mean-field result at zero temperature for one condensate
with its chemical potential: with a second condensate that has a chemical
potential of its own the integrals are hyperelliptic.  The state that it
describes is a local minimum, since at gaps of the order of $\lambda$ the
potential turns down.  Where it applies, no momentum integral is left to be
computed: the potential, the gap equation and the density are closed
expressions in the mass, the chemical potential and the gap, and the scale
that the integral does not fix is a property of the pair in the vacuum.

\begin{acknowledgments}
I am grateful to Prof.~Zhukovsky, Prof.~Klimenko, A.~Akhmeteli,
Academician~Kvinikhidze and Dr.~Zhokhov.

This work was supported by the Shota Rustaveli National Science Foundation of
Georgia (SRNSFG) under grant no.~FR-25-7544, ``Nonperturbative Analysis of the
QCD Critical Point: Regularization problems in the Effective Models and
Renormalization Group Techniques''.

During the preparation of the manuscript, Claude (Anthropic, Claude Fable 5.2)
helped to write and edit the text and to assemble the Supplemental Material,
together with the script that generates and checks it.  The author reviewed
and revised all AI-assisted text and takes full responsibility for the
scientific content and final wording.
\end{acknowledgments}

\appendix

\section{The response of the figure}\label{app:figure}

The quadratic \eqref{eq:quad} contains $S^{2}$, $E_p^{2}$, $\mu^{2}$ and
$\Delta^{2}$.  Differentiated with respect to all four, with \eqref{eq:prodheights} for the
coefficient of $d(S^{2})$, it reads
\begin{equation}\label{eq:diffrel}
  E^{+}E^{-}\,d\bigl(S^{2}\bigr)=4h_\mu^{2}\,d\bigl(E_p^{2}\bigr)+4h_E^{2}\,d\bigl(\mu^{2}\bigr)+S^{2}\,d\bigl(\Delta^{2}\bigr) .
\end{equation}
The height above $\mu$ goes with the point $E_p$, the height above $E_p$ with
the point $\mu$, and the radius with the height of the apex.  With
$E_p^{2}=p^{2}+m^{2}$ these are the derivatives of the sum $S$,
\begin{equation}\label{eq:derivs}
  \frac{dS}{dp}=\frac{4p\,h_\mu^{2}}{E^{+}E^{-}S},\qquad
  \frac{\partial S}{\partial m}=\frac{4m\,h_\mu^{2}}{E^{+}E^{-}S},\qquad
  \frac{\partial S}{\partial\mu}=\frac{4\mu\,h_E^{2}}{E^{+}E^{-}S},\qquad
  \frac{\partial S}{\partial\Delta}=\frac{\Delta\,S}{E^{+}E^{-}} .
\end{equation}
The Bogoliubov angles of the two branches at the momentum $p$ and the
amplitudes are
\begin{equation}\label{eq:bogangles}
  \cos\theta_\pm(p)=\frac{E_p\pm\mu}{E^{\pm}},\qquad
  \sin\theta_\pm(p)=\frac{\Delta}{E^{\pm}},\qquad
  u_\pm=\cos\frac{\theta_\pm}{2},\quad v_\pm=\sin\frac{\theta_\pm}{2} ,
\end{equation}
and with them the three contents \eqref{eq:modes} are
$\tfrac12\partial S/\partial\mu$, $\tfrac12\partial S/\partial E_p$ and
$\tfrac12\partial S/\partial\Delta$.

\section{The path and the integration by parts}\label{app:path}

At $p=0$ the two heights and their relations \eqref{eq:heights} and
\eqref{eq:prodheights} are
\begin{equation}\label{eq:h0}
  \begin{gathered}
  h_m^{2}=\tfrac14(\ell_++\ell_-)^{2}-m^{2},\qquad
  h_{\mu0}^{2}=\tfrac14(\ell_++\ell_-)^{2}-\mu^{2},\\
  h_m h_{\mu0}=\tfrac12(\ell_++\ell_-)\,\Delta,\qquad
  h_m^{2}+h_{\mu0}^{2}-\Delta^{2}=\ell_+\ell_- .
  \end{gathered}
\end{equation}
Eliminating $S$ and $h_\mu$ from \eqref{eq:ell2} and \eqref{eq:heights} gives
the momentum as a function of the bounded height,
\begin{equation}\label{eq:p2beta}
  p^{2}=\frac{\bigl(h_m^{2}-h_E^{2}\bigr)\bigl(h_E^{2}+h_{\mu0}^{2}-\Delta^{2}\bigr)}{h_E^{2}-\Delta^{2}} ,
\end{equation}
a ratio in which the substitution \eqref{eq:betaphi} turns the two factors of
the numerator into $h_m^{2}\sin^{2}\varphi$ and
$\ell_+\ell_-(1-k^{2}\sin^{2}\varphi)$.  Along the path, with
$s\equiv\sin^{2}\varphi_1-\sin^{2}\varphi$,
\begin{equation}\label{eq:onpath}
  S=\frac{2\mu\cos\varphi}{\sqrt s},\qquad
  p=\sqrt{\ell_+\ell_-}\;\frac{\sin\varphi\,\sqrt{1-k^{2}\sin^{2}\varphi}}{\sqrt s},\qquad
  h_\mu=\frac{\mu\cos\varphi_1}{\sqrt s} .
\end{equation}
The step \eqref{eq:step} follows from these and \eqref{eq:prodheights}, and
\eqref{eq:sLam} from these and \eqref{eq:p2h}.

For the integration by parts, the integrand $p^{2}S$ is even in $p$ and grows
as $2p^{3}$, so we try the odd function $W=p\,S\,(w_1p^{2}+w_2+w_3S^{2})$ and
require that $p^{2}S-dW/dp$ be
$(c_1+c_2h_E^{2}+c_3h_\mu^{2})/(E^{+}E^{-}S)$.  With $dS/dp$ of
\eqref{eq:derivs},
\begin{equation}\label{eq:dW}
  E^{+}E^{-}S\,\frac{dW}{dp}=\bigl(E^{+}E^{-}S^{2}+4p^{2}h_\mu^{2}\bigr)\,Q
  +p^{2}S^{2}\bigl(2w_1E^{+}E^{-}+8w_3h_\mu^{2}\bigr),\qquad Q=w_1p^{2}+w_2+w_3S^{2},
\end{equation}
a polynomial in $p^{2}$, $S^{2}$ and the parameters once $E^{+}E^{-}$ and
$h_\mu^{2}$ are expressed in terms of $S^{2}$ by \eqref{eq:prodheights}.  The
requirement, with the quadratic \eqref{eq:quad} used to remove $S^{4}$, is a
linear system for the six numbers, and it has one solution:
\begin{equation}\label{eq:Wshort}
  W=\frac{p\,S\,\bigl(16\,p^{2}+10\,\rho^{2}-S^{2}\bigr)}{48} ,
\end{equation}
\begin{equation}\label{eq:coefs}
  c_1=\frac{\mu^{2}\Delta^{2}}{6}\bigl(12\mu^{2}-11m^{2}-3\Delta^{2}\bigr),\qquad
  c_2=\frac{\mu^{2}}{6}\bigl(2\mu^{2}-5m^{2}-13\Delta^{2}\bigr),\qquad
  c_3=2\mu^{2}\Delta^{2}-\frac{\rho^{4}}{2} .
\end{equation}

The sums of the three weights up to $\Lambda$, taken with the step
\eqref{eq:step}, are
\begin{equation}\label{eq:Tsums}
  \begin{alignedat}{2}
  T_1&\equiv\int_0^{\Lambda}\frac{dp}{E^{+}E^{-}S}&&=g\,\EllF(\varphi_\Lambda,k),\\[8pt]
  T_2&\equiv\int_0^{\Lambda}\frac{h_E^{2}\,dp}{E^{+}E^{-}S}&&
     =g\bigl[\ell_+\ell_-\,\EllE(\varphi_\Lambda,k)-(h_{\mu0}^{2}-\Delta^{2})\,\EllF(\varphi_\Lambda,k)\bigr],\\[8pt]
  T_3&\equiv\int_0^{\Lambda}\frac{h_\mu^{2}\,dp}{E^{+}E^{-}S}&&=g\,h_{\mu0}^{2}\,\Pi(\varphi_\Lambda;\alpha^{2},k),
  \end{alignedat}
\end{equation}
with $g\equiv1/(2\mu\sqrt{\ell_+\ell_-})$, and the identity
\eqref{eq:oneline} integrates to
\begin{equation}\label{eq:physclosed}
  I(\Lambda)=W(\Lambda)+c_1\,T_1+c_2\,T_2+c_3\,T_3 ,
\end{equation}
which is \eqref{eq:closed} term by term.

\section{The end of the arc}\label{app:end}

What grows with $\Lambda$ is decided at large momenta, and there only the
height above $\mu$ grows: by \eqref{eq:p2h},
$h_\mu^{2}=\Lambda^{2}+\rho^{2}-\mu^{2}+O(\Lambda^{-2})$ at $p=\Lambda$, while
$h_E$ comes down to $\Delta$.  In terms of $h_\mu$ the whole figure is
rational:
\begin{equation}\label{eq:inheight}
  \begin{gathered}
  S^{2}=4\bigl(h_\mu^{2}+\mu^{2}\bigr),\qquad
  h_E^{2}=\Delta^{2}+\frac{\mu^{2}\Delta^{2}}{h_\mu^{2}},\qquad
  E^{+}E^{-}=h_\mu^{2}+\frac{\mu^{2}\Delta^{2}}{h_\mu^{2}},\\[4pt]
  p^{2}=h_\mu^{2}+\mu^{2}-\rho^{2}-\frac{\mu^{2}\Delta^{2}}{h_\mu^{2}},\qquad
  p\,dp=\frac{E^{+}E^{-}}{h_\mu}\,dh_\mu,\\[4pt]
  W=\frac{p\,S}{48}\Bigl(12h_\mu^{2}+12\mu^{2}-6\rho^{2}-\frac{16\mu^{2}\Delta^{2}}{h_\mu^{2}}\Bigr)
  =\frac{h_\mu^{4}}{2}+\frac{(2\mu^{2}-\rho^{2})\,h_\mu^{2}}{2}+w_0+O(h_\mu^{-2}),\\[4pt]
  w_0=\frac{(2\mu^{2}-\rho^{2})^{2}}{8}-\frac{11}{12}\,\mu^{2}\Delta^{2}-\frac{\rho^{4}}{16},\\[4pt]
  \frac{c_1+c_2\,h_E^{2}+c_3\,h_\mu^{2}}{E^{+}E^{-}\,S}\;dp
  =\frac{c_1+c_2\,h_E^{2}+c_3\,h_\mu^{2}}{p\,S}\;\frac{dh_\mu}{h_\mu}
  =\frac{c_3}{2}\,\frac{dh_\mu}{h_\mu}\,\bigl[1+O(h_\mu^{-2})\bigr] .
  \end{gathered}
\end{equation}
The boundary term leaves two powers of the height and the constant $w_0$,
which give \eqref{eq:WLam}.  In the last line the numerators of the first two
kinds stay finite, so these fall as $h_\mu^{-3}$ and enter the finite part
only, and the third kind leaves $\tfrac12c_3$ for every unit of
$\ln h_\mu$.

By \eqref{eq:prodheights} the third weight is not independent of the other
two,
\begin{equation}\label{eq:thirdJ}
  \frac{h_\mu^{2}}{E^{+}E^{-}S}=\frac1S-\frac{h_E^{2}-\Delta^{2}}{E^{+}E^{-}S} ,
\end{equation}
so the third kind enters only through the integral of $1/S$ over the
momentum:
\begin{equation}\label{eq:Jexact}
  \begin{gathered}
  J(\Lambda)\equiv\int_0^{\Lambda}\frac{dp}{E^{+}+E^{-}}
  =\frac12\,\operatorname{artanh}t_\Lambda
  +\frac{\ell_+\ell_-\,\EllE(\varphi_\Lambda,k)-h_{\mu0}^{2}\,\Pi(\varphi_\Lambda;\omega^{2},k)}{2\mu\sqrt{\ell_+\ell_-}},\\
  t_\Lambda=\frac{2\,p\,h_{\mu0}^{2}}{S\,\bigl(h_E^{2}+h_{\mu0}^{2}-\Delta^{2}\bigr)}\bigg|_{p=\Lambda},\qquad
  \omega^{2}=k^{2}\sin^{2}\varphi_1=\frac{4\mu^{2}-(\ell_+-\ell_-)^{2}}{4\,\ell_+\ell_-} .
  \end{gathered}
\end{equation}
Both sides vanish at $\Lambda=0$, and the derivative of the right side is a
rational function on the arc, equal to $1/S$.  In the tables this is the
change of parameter of the third kind~\cite{DLMF}, which moves the pole of
$\Pi$ away from the end of the arc at the price of an elementary function.
Since $\omega^{2}\sin^{2}\varphi_1<1$, the integral
$\Pi(\varphi;\omega^{2},k)$ is finite up to the end of the arc, and the
logarithm is the $\operatorname{artanh}$: $t_\Lambda$ grows from $0$ to $1$,
with $1-t_\Lambda^{2}=(\ell_+\ell_-+\mu^{2})/\Lambda^{2}+O(\Lambda^{-4})$, so
that
\begin{equation}\label{eq:Jinf}
  J(\Lambda)=\frac12\ln\frac{2\Lambda}{\sqrt{\ell_+\ell_-+\mu^{2}}}
  +\frac{\ell_+\ell_-\,\EllE(\varphi_1,k)-h_{\mu0}^{2}\,\Pi(\varphi_1;\omega^{2},k)}{2\mu\sqrt{\ell_+\ell_-}}
  +O(\Lambda^{-2}) .
\end{equation}

With \eqref{eq:thirdJ} and \eqref{eq:Jexact} the integral \eqref{eq:closed}
takes the form
\begin{equation}\label{eq:closedJ}
  \begin{split}
  I(\Lambda)=W(\Lambda)+\frac{c_3}{2}\operatorname{artanh}t_\Lambda
  +\frac{1}{2\mu\sqrt{\ell_+\ell_-}}\Bigl[&
  \bigl(c_1-c_2\,(h_{\mu0}^{2}-\Delta^{2})+c_3\,h_{\mu0}^{2}\bigr)\,\EllF(\varphi_\Lambda,k)
  +c_2\,\ell_+\ell_-\,\EllE(\varphi_\Lambda,k)\\
  &-c_3\,h_{\mu0}^{2}\,\Pi(\varphi_\Lambda;\omega^{2},k)\Bigr] ,
  \end{split}
\end{equation}
still exact at every $\Lambda$.  The powers are in $W$, the logarithm is the
$\operatorname{artanh}$, and the bracket has a limit, which gives
\eqref{eq:asymp} and \eqref{eq:Cfin}.  The same finite part is a convergent
integral over all momenta,
\begin{equation}\label{eq:Cint}
  \mathcal C(m,\mu,\Delta)=\int_0^{\infty}\Bigl[p^{2}\bigl(E^{+}+E^{-}\bigr)-2p^{3}-\rho^{2}p
  -\frac{c_3/2}{\sqrt{p^{2}+\ell_+\ell_-+\mu^{2}}}\Bigr]dp .
\end{equation}

\section{The constants of the crossover}\label{app:constants}

By \eqref{eq:modes} each of the three contents is a combination of the
weights $1$, $h_E^{2}$ and $h_\mu^{2}$ over the step, and the factor $p^{2}$
can be taken away as in Appendix~\ref{app:path}.  With a boundary term
$p\,S\,(b_1+b_2/h_\mu^{2}+b_3/h_\mu^{4})$ and numbers $b_i$ to be found, the
same search gives
\begin{equation}\label{eq:p2weights}
  \begin{aligned}
  p^{2}&=h_\mu^{2}-h_E^{2}+\mu^{2}-m^{2},\\
  \frac{p^{2}\,h_E^{2}}{E^{+}E^{-}S}&=\frac{d}{dp}\Bigl[-\frac{\Delta^{2}}{12}\,\frac{p\,S}{h_\mu^{2}}\Bigr]
  +\frac{\Delta^{2}\bigl(\mu^{2}-\tfrac13m^{2}\bigr)+\tfrac13\bigl(\mu^{2}-m^{2}-2\Delta^{2}\bigr)h_E^{2}+\Delta^{2}h_\mu^{2}}{E^{+}E^{-}S},\\
  \frac{p^{2}\,h_\mu^{2}}{E^{+}E^{-}S}&=\frac{d}{dp}\Bigl[\frac{p\,S}{8}\Bigr]
  -\frac{\mu^{2}\Delta^{2}+\mu^{2}h_E^{2}+\rho^{2}h_\mu^{2}}{2\,E^{+}E^{-}S} .
  \end{aligned}
\end{equation}
The first relation needs no boundary term; it is the second of
\eqref{eq:prodheights}.  With the sums $T_1$, $T_2$, $T_3$ of
\eqref{eq:Tsums} and the third expressed through the function $J$ of
\eqref{eq:Jexact},
\begin{equation}\label{eq:Tdef}
  T_3=\int_0^{\Lambda}\!\frac{h_\mu^{2}\,dp}{E^{+}E^{-}S}=J(\Lambda)-T_2+\Delta^{2}T_1 ,
\end{equation}
the three contents are
\begin{equation}\label{eq:Qclosed}
  \begin{aligned}
  Q_\mu&=2\mu\Bigl[-\frac{\Delta^{2}}{12}\,\frac{\Lambda S}{h_\mu^{2}}
  +\Delta^{2}\bigl(\mu^{2}-\tfrac13m^{2}\bigr)T_1+\tfrac13\bigl(\mu^{2}-m^{2}-2\Delta^{2}\bigr)T_2+\Delta^{2}\,T_3\Bigr],\\
  Q_m&=2m\Bigl[\frac{\Lambda S}{8}-\tfrac12\mu^{2}\Delta^{2}\,T_1-\tfrac12\mu^{2}\,T_2-\tfrac12\rho^{2}\,T_3\Bigr],\\
  Q_\Delta&=2\Delta\Bigl[\frac{\Lambda S}{8}+\mu^{2}\bigl(\mu^{2}-m^{2}-\tfrac12\Delta^{2}\bigr)T_1-\tfrac32\mu^{2}\,T_2
  +\bigl(\mu^{2}-\tfrac12\rho^{2}\bigr)T_3\Bigr],
  \end{aligned}
\end{equation}
with $S$ and $h_\mu$ taken at $p=\Lambda$, exact at every $\Lambda$.

The sums $T_1$ and $T_2$ converge, and $T_3$ grows with $J$, as
$\tfrac12\ln(2\Lambda/\ell_0)$ by \eqref{eq:Jinf}.  The coefficient of the
logarithm in each content is therefore half the coefficient of $T_3$:
$\mu\Delta^{2}$ in $Q_\mu$, and $\mu^{2}-\tfrac12\rho^{2}$ in
$Q_\Delta/\Delta$.  In $V$ the logarithm has the coefficient
$\tfrac12m^{2}-q_B^{2}$,
\begin{equation}\label{eq:Vinf}
  V(\Lambda)=\tfrac12\Lambda^{2}+\tfrac14m^{2}+\bigl(\tfrac12m^{2}-q_B^{2}\bigr)\ln\frac{2\Lambda}{m}
  -q_B\sqrt{m^{2}-q_B^{2}}\,\arccos\frac{q_B}{m}+O(\Lambda^{-2}) ,
\end{equation}
and the difference of the two coefficients,
$\mu^{2}-\tfrac14m_B^{2}-\tfrac12\Delta^{2}$, is the coefficient of the tail
of $f_D$ in \eqref{eq:tails}.  The boundary term of $Q_\Delta/\Delta$ is
$\tfrac14\Lambda S=\tfrac12\Lambda^{2}+\tfrac14\rho^{2}+O(\Lambda^{-2})$, and
its $\tfrac12\Lambda^{2}$ is the part of $V$ that the coupling took; that of
$Q_\mu$ tends to $-\tfrac13\mu\Delta^{2}$.  At a large upper limit the density
and the denominator are thus $\mu\Delta^{2}\ln(2\Lambda/\ell_0)+q_\mu$ and
$(\mu^{2}-\tfrac14m_B^{2}-\tfrac12\Delta^{2})\ln(2\Lambda/\ell_0)+D_0$, up to
terms of the order $\Lambda^{-2}$, with
\begin{equation}\label{eq:qmuD0}
  \begin{aligned}
  q_\mu&=2\mu\Bigl[-\frac{\Delta^{2}}{6}+\Delta^{2}\bigl(\mu^{2}-\tfrac13m^{2}+\Delta^{2}\bigr)T_1^{\infty}
  +\tfrac13\bigl(\mu^{2}-m^{2}-5\Delta^{2}\bigr)T_2^{\infty}+\Delta^{2}J_1\Bigr],\\
  D_0&=\frac{\Delta^{2}}{4}+\bigl(q_B^{2}-\tfrac12m^{2}\bigr)\ln\frac{\ell_0}{m}
  +q_B\sqrt{m^{2}-q_B^{2}}\,\arccos\frac{q_B}{m}\\
  &\quad+2\Bigl[\bigl(\mu^{2}(\mu^{2}-m^{2}-\tfrac12\Delta^{2})+\Delta^{2}(\mu^{2}-\tfrac12\rho^{2})\bigr)T_1^{\infty}
  -\bigl(\tfrac52\mu^{2}-\tfrac12\rho^{2}\bigr)T_2^{\infty}+\bigl(\mu^{2}-\tfrac12\rho^{2}\bigr)J_1\Bigr],
  \end{aligned}
\end{equation}
where $T_1^{\infty}$ and $T_2^{\infty}$ are $T_1$ and $T_2$ at the end of the
arc, $\varphi_\Lambda\to\varphi_1$, and $J_1$ is the second term of
\eqref{eq:Jinf}.  With all momenta integrated and the logarithm measured from
$\lambda$, these are \eqref{eq:Qinf}, and nothing is dropped.  In $D_0$ the
term with $\ln(\ell_0/m)$ moves the logarithm of $V$ from the energy $m$ to
the energy $\ell_0$.

The two equations are the derivatives of one potential.  With all momenta
integrated the integral is $I_\lambda$ of \eqref{eq:main}, and the logarithm of
the loop \eqref{eq:Vdef} is measured from the same scale,
\begin{equation*}
  V_\lambda\equiv\lim_{\Lambda\to\infty}\Bigl[V(\Lambda)-\frac{\Lambda^{2}}{2}
  -\bigl(\tfrac12m^{2}-q_B^{2}\bigr)\ln\frac{\Lambda}{\lambda}\Bigr]
  =\frac{m^{2}}{4}+\bigl(\tfrac12m^{2}-q_B^{2}\bigr)\ln\frac{2\lambda}{m}
  -q_B\sqrt{m^{2}-q_B^{2}}\,\arccos\frac{q_B}{m} .
\end{equation*}
The power taken out of $V$ is the one that the coupling took in
\eqref{eq:split}.  Up to a constant the potential is then
$\Omega/N=\Delta^{2}V_\lambda-I_\lambda$, and its derivatives are
$D=-\partial(\Omega/N)/\partial(\Delta^{2})$, which is
$\partial I_\lambda/\partial(\Delta^{2})-V_\lambda$, and
$Q_\mu=-\tfrac12\partial(\Omega/N)/\partial\mu=\tfrac12\partial I_\lambda/\partial\mu$.
At a fixed mass and a fixed binding of the pair the potential depends on the
scale through
\begin{equation}\label{eq:runB}
  \lambda\frac{\partial}{\partial\lambda}\,\frac{\Omega}{N}=\frac{m^{4}}{4}+\Omega_B ,
\end{equation}
that is, the potential contains $\ln\lambda$ with the coefficient
$\tfrac14m^{4}+\Omega_B$.  The first term belongs to the vacuum, and the
second is the only function of the state that the logarithm of the scale
multiplies.

\section{The nonrelativistic corner}\label{app:corner}

Let $m\to\infty$ with $\mu_{\rm nr}\equiv\mu-m$ and $\Delta$ fixed, the
momenta and the upper limit being of the order $\sqrt{2m\Delta}$, so that
$p^{2}/2m$ stays finite, and let
$\xi_p\equiv E_p-\mu$ be the energy of the free particle measured from the
chemical potential, so that $E^{-}=\sqrt{\xi_p^{2}+\Delta^{2}}$ exactly.  By
\eqref{eq:prodheights} the two heights are, exactly,
\begin{equation}\label{eq:nrexact}
  h_\mu^{2}-h_E^{2}=\xi_p\,(\xi_p+2\mu),\qquad
  h_\mu^{2}+h_E^{2}=E^{+}E^{-}+\Delta^{2} .
\end{equation}
In the corner $\xi_p\to p^{2}/2m-\mu_{\rm nr}$ stays finite while
$E^{+}\to2m$, and with the amplitudes of BCS theory,
$u_p^{2},v_p^{2}=\tfrac12(1\pm\xi_p/E^{-})$,
\begin{equation}\label{eq:nrheights}
  h_E^{2}=2mE^{-}\,v_p^{2}\,\bigl[1+O(m^{-1})\bigr],\qquad
  h_\mu^{2}=2mE^{-}\,u_p^{2}\,\bigl[1+O(m^{-1})\bigr],\qquad
  \frac{dp}{E^{+}E^{-}S}=\frac{dp}{4m^{2}E^{-}}\,\bigl[1+O(m^{-1})\bigr] .
\end{equation}
The square of the height above $E_p$ becomes the occupation of the mode, and
the square of the height above $\mu$ its complement.  The three sums
\eqref{eq:Tsums} become
\begin{equation}\label{eq:nrsums}
  T_1\to\frac{1}{4m^{2}}\int_0^{\Lambda}\!\frac{dp}{E^{-}},\qquad
  T_2\to\frac{1}{2m}\int_0^{\Lambda}\!v_p^{2}\,dp,\qquad
  T_3\to\frac{\Lambda}{2m}-\frac{1}{2m}\int_0^{\Lambda}\!v_p^{2}\,dp .
\end{equation}
The third sum is $\Lambda/2m$ minus the second: with the level of the
antiparticle frozen, the third weight brings no new function.

The triangle opens into an angle: the apex sees the near end of the base at
the distance $\ell_-=\sqrt{\mu_{\rm nr}^{2}+\Delta^{2}}$ and the far end along
the axis.  Then
\begin{equation}\label{eq:nrmod}
  k^{2},\ \omega^{2}\ \longrightarrow\ \kappa^{2}=\frac12\Bigl(1+\frac{\mu_{\rm nr}}{\sqrt{\mu_{\rm nr}^{2}+\Delta^{2}}}\Bigr),
  \qquad
  \cos^{2}\varphi_1=\frac{\ell_--\mu_{\rm nr}}{m}+O(m^{-2}),\qquad
  \alpha^{2}\ \longrightarrow\ 1 .
\end{equation}
The amplitude of the end tends to $\pi/2$: the arc grows to a quarter of the
ellipse, and the integrals become complete.  The parameter of the third kind
tends to unity in \eqref{eq:closed} and to the modulus in \eqref{eq:closedJ}, and at
those values the integral of the third kind is not a new function:
\begin{equation}\label{eq:Pi1}
  \begin{gathered}
  \Pi(\varphi;1,k)=\EllF(\varphi,k)-\frac{\EllE(\varphi,k)-\tan\varphi\,\sqrt{1-k^{2}\sin^{2}\varphi}}{1-k^{2}},\\
  \Pi(\varphi;k^{2},k)=\frac{1}{1-k^{2}}\Bigl[\EllE(\varphi,k)-\frac{k^{2}\sin\varphi\cos\varphi}{\sqrt{1-k^{2}\sin^{2}\varphi}}\Bigr] .
  \end{gathered}
\end{equation}
What is left is $\EllK(\kappa)$ and $\EllE(\kappa)$, the functions and the
modulus of Ref.~\cite{MariniPistolesiStrinati1998}.  One of their integrals
can be compared directly.  None of the coefficients of the three growing
terms of \eqref{eq:asymp} contains $m$ and $\mu$ together, so the mixed
derivative of
the integral converges by itself, and in the corner
\begin{equation}\label{eq:nrI5}
  \begin{gathered}
  -\frac{\partial^{2}}{\partial m\,\partial\mu}\int_0^{\infty}p^{2}\bigl(E^{+}+E^{-}\bigr)\,dp
  =\int_0^{\infty}\frac{m\Delta^{2}}{E_p}\Bigl[\frac{1}{(E^{-})^{3}}-\frac{1}{(E^{+})^{3}}\Bigr]p^{2}\,dp
  \ \longrightarrow\ (2m)^{3/2}\,\Delta^{1/2}\,I_5(x_0),\\
  I_5(x_0)\equiv\int_0^{\infty}\frac{x^{2}\,dx}{\bigl[(x^{2}-x_0)^{2}+1\bigr]^{3/2}}
  =(1+x_0^{2})^{1/4}\,\EllE(\kappa)-\frac{\EllK(\kappa)}{4x_1^{2}\,(1+x_0^{2})^{1/4}},\\
  x_0=\frac{\mu_{\rm nr}}{\Delta},\qquad x_1^{2}=\frac{\sqrt{1+x_0^{2}}+x_0}{2} ,
  \end{gathered}
\end{equation}
with $p^{2}=2m\Delta\,x^{2}$; on the left the derivative is taken under the
integral.  The relative difference of the two sides falls as $1/m$.

For the two equations of Sec.~\ref{sec:crossover} we measure the momentum in
the units of the nonrelativistic problem, $p=\sqrt{2m\Delta}\,x$, keep the
upper limit and $q_B$ fixed in these units,
$\Lambda=\sqrt{2m\Delta}\,x_\Lambda$ and $q_B=\sqrt{2m\Delta}\,\hat q$, and let
$m\to\infty$.  Then, with $\xi_x=x^{2}-x_0$ and $E_x=\sqrt{\xi_x^{2}+1}$,
\begin{equation}\label{eq:nreqs}
  \begin{gathered}
  \frac{D(\Lambda)}{m\sqrt{2m\Delta}}\ \longrightarrow\
  \int_0^{x_\Lambda}\Bigl(\frac{x^{2}}{E_x}-1\Bigr)dx+\hat q\,\Bigl[\frac{\pi}{2}-\arctan\frac{\hat q}{x_\Lambda}\Bigr],\\
  \frac{Q_\mu(\Lambda)}{\tfrac12(2m\Delta)^{3/2}}\ \longrightarrow\
  \int_0^{x_\Lambda}x^{2}\Bigl(1-\frac{\xi_x}{E_x}\Bigr)dx ,
  \end{gathered}
\end{equation}
with corrections of the order $\Delta/m$ at fixed $x_\Lambda$.  As
$x_\Lambda\to\infty$ the right sides are $I_1(x_0)+\tfrac12\pi\hat q$ and
$I_2(x_0)$, with the two integrals of
Ref.~\cite{MariniPistolesiStrinati1998}, its Eqs.~(2.9) and (2.10), and the
gap equation $D=0$ and the density $n=Q_\mu/\pi^{2}$ of two spin states are
its Eqs.~(2.8a) and (2.8b), with the scattering length $a=1/q_B$.  Both
signs of $q_B$ are needed: $I_1(x_0)$ vanishes at $x_0=0.8604$, the point
$1/a=0$, and for larger $x_0$, on the side of BCS pairing, $q_B<0$.

If the upper limit is of the order of $m$ or above it, as when all momenta
are integrated and the logarithm is measured from $\lambda$, the region
$p\sim m$ contributes as well, and the logarithm of \eqref{eq:Qinf} does not
vanish identically in the corner.  In the gap equation its coefficient is
$\mu^{2}-\tfrac14m_B^{2}-\tfrac12\Delta^{2}\simeq2m\Delta\,(x_0+\hat q^{2})$,
against the size $m\sqrt{2m\Delta}$ of the equation, and in the density it is
$\mu\Delta^{2}\simeq m\Delta^{2}$, against $\tfrac12(2m\Delta)^{3/2}$.  In both
the corrections are of the relative order $\sqrt{\Delta/m}\,\ln(2\lambda/m)$,
that is, of the order $k_Fr_e$.

\section{How the formulas were checked}\label{app:checks}

Every numbered formula of this paper that is not a definition is verified by
the script that generates the Supplemental Material~\cite{supp}, symbolically
wherever the statement is algebraic; the list of the formulas, by their
numbers here, with the check that closes each, is part of that material.  The
identities of the path are checked in the field of the two
numbers $\sin^{2}\varphi_1$ and $\cosh^{2}\xi_1$, in which every quantity is
rational and no square root has to be chosen.  The other statements are tested
numerically to thirty digits: the integral at an upper limit,
\eqref{eq:closed} and \eqref{eq:closedJ}, against a quadrature; the finite
part \eqref{eq:Cfin} against the convergent integral \eqref{eq:Cint}, and the
remainder of \eqref{eq:asymp} at three upper limits; two windows of
\eqref{eq:anyrule}; the three contents \eqref{eq:Qclosed}, and how they and
the potential approach \eqref{eq:Qinf} and \eqref{eq:runB} as the upper limit
grows; the effective range, \eqref{eq:Dexp}--\eqref{eq:Zfield}, and the limit
to the two-channel model \eqref{eq:twochannel}; the nonrelativistic corner at
three masses; and, for two apexes, the integration by parts behind
\eqref{eq:survn}.

Three checks are against results obtained elsewhere.  Reference~\cite{companion}
gives the closed form of the same integral in three forms, obtained there by
Hermite's reduction, and they agree with \eqref{eq:closed},
\eqref{eq:closedJ} and \eqref{eq:Cfin} to the working precision; the finite
part of that reference is $\mathcal C+\tfrac12c_3\ln(2/\ell_0)$ in the
notation used here.  In the three limits in which the integral is elementary,
$\mu\to0$ and $\Delta\to0$ on either side of $\mu=m$, the finite part tends
to the elementary values.  In the nonrelativistic corner the two equations
become those of Ref.~\cite{MariniPistolesiStrinati1998}
(Appendix~\ref{app:corner}).  The script, its output and the code of the
figures are part of the Supplemental Material.

\bibliography{refs}
\end{document}